\documentclass[10pt,journal,compsoc]{IEEEtran}

\usepackage{cite}
\usepackage{amsmath,amssymb,amsfonts}
\usepackage{amsthm}
\usepackage{algorithmic}
\usepackage[ruled,vlined,linesnumbered,resetcount]{algorithm2e}
\usepackage{graphicx}
\usepackage{subfigure}
\usepackage{multirow}
\usepackage{makecell}
\usepackage{array}
\usepackage{textcomp}
\usepackage{xcolor}
\usepackage{comment}
\usepackage{color}
\usepackage{pgfplots}
\usepackage{tikz} 
\usepackage{eurosym}
\usepackage{filecontents}

\theoremstyle{definition}
\newtheorem{definition}{Definition}[section]
\newtheorem{theorem}{Theorem}

\begin{document}

\title{Trajectory-Based Spatiotemporal Entity Linking}

\author{Fengmei~Jin, Wen~Hua, Thomas~Zhou, Jiajie~Xu, Matteo~Francia, Maria~E~Orlowska, Xiaofang~Zhou,~\IEEEmembership{Fellow,~IEEE}
}

\IEEEtitleabstractindextext{
\begin{abstract}
Trajectory-based spatiotemporal entity linking is to match the same moving object in different datasets based on their movement traces. It is a fundamental step to support spatiotemporal data integration and analysis. In this paper, we study the problem of spatiotemporal entity linking using effective and concise signatures extracted from their  trajectories. This linking problem is formalized as a $k$-nearest neighbor ($k$-NN) query on the signatures. Four representation strategies (sequential, temporal, spatial, and spatiotemporal) and two quantitative criteria (commonality and unicity) are investigated for signature construction. A simple yet effective dimension reduction strategy is developed together with a novel indexing structure called the WR-tree to speed up the search. A number of optimization methods are proposed to improve the accuracy and robustness of the linking. Our extensive experiments on real-world datasets verify the superiority of our approach over the state-of-the-art solutions in terms of both accuracy and efficiency.

\end{abstract}

}

\maketitle

\IEEEdisplaynontitleabstractindextext

% \fi
\IEEEpeerreviewmaketitle

\section{Introduction}
\label{sec:introduction}
With the prevalence of location-capturing devices and location-based services comes an ever-increasing amount and variety of spatial trajectory data, such as vehicle trajectories, sensor readings, telecom tokens, and location check-ins. The mining of mobility patterns has become an important research topic due to its utility in significant real-world applications including transport management and urban planning. One interesting problem to study is the extent to which individual movements are unique and distinctive, i.e., the possibility of identifying an individual based on her movement patterns extracted from the historical traces. Such study can benefit various real-world applications.

\noindent\textbf{Scenario 1:} A user may have multiple social network accounts. By matching accounts across platforms (e.g., Twitter and Facebook), a more comprehensive user profile can be constructed, which potentially improves the performance of personalized recommendation.

\noindent\textbf{Scenario 2:} User monitoring is possible by linking phone numbers used by a single person. This is extremely important for security control such as criminal tracking.

\noindent\textbf{Scenario 3:} A taxi driver can register with multiple 
Uber-like companies. Once this information is combined together through entity linking, different driving behaviors of the same user could be observed, stimulating transformation of operation patterns in the company.

A recent study on mobile phone users \cite{de2013unique} has shown that the uniqueness of human mobility is high enough that 4 randomly-sampled spatiotemporal points could uniquely identify 95\% of individuals. However, can the same method be utilized to identify other types of moving objects? We conducted a preliminary experimental study on a real-world taxi dataset, and observed that only 15\% of taxis could be identified using the method proposed in \cite{de2013unique}. This motivated us to explore a better representation of mobility patterns for accurate moving object linking. Unlike traditional methods which discover mobility patterns to capture human's collective \cite{Li2010Swarm}\cite{Zheng2013Gather}, sequential \cite{Cao2005}\cite{Giannotti2007} or periodic \cite{Cao2007}\cite{Li2010} movement behaviors, in this work, we aim at extracting patterns, hereafter referred to as \textit{signatures}, for entity linking, i.e., matching traces made by the same entity in different datasets such as social media users and taxi drivers. 

\subsection{Challenges and Contributions}
\textit{How to effectively represent and quantify the signature of a moving object to guarantee the accuracy of object linking?} 
Trajectories are intrinsically spatial, temporal, and sequential. Hence, the object signature should also capture these characteristics. Existing trajectory pattern mining algorithms \cite{Cao2005}\cite{Giannotti2007}\cite{Cao2007}\cite{Li2010} extract sequential and temporal patterns from trajectories, but ignore spatial features. However, our empirical study on a real-world taxi dataset shows that spatial patterns are the most effective for identifying individuals, as opposed to the other two features. Additionally, signatures need to encode patterns quantitatively to allow for similarity calculations, which is a non-trivial task. Intuitively, the quantitative signature should be able to encapsulate behavior that is not only ubiquitous throughout the whole trace of the individual but also highly discriminative from one individual to another. In this paper, two metrics are proposed to quantify a signature from an individual's historical trace: 1) \textit{commonality}, the frequency of the signature in the individual's trace, and 2) \textit{unicity}, the extent to which an individual can be distinguished from others by their signatures.

\textit{How to improve the efficiency of moving object linking?} 
This work reduces the task of object linking to a $k$-nearest neighbor ($k$-NN) search problem: given a query object, we search in a set of moving objects to find the top-$k$ candidates based on signature similarity. The most straightforward approach is a linear scan which calculates all pairwise similarities one by one. However, this is infeasible in practice due to the following challenges:

\textit{Curse of dimensionality:} 
The signatures can be extremely large. In the worst case, when every point in an individual's historical trace contributes to her personalized profile, the size of signatures could be as large as the number of distinct points in the entire dataset (or even larger if sequential patterns are considered). The computational complexity of comparing two signatures is $O(d)$ where $d$ represents the dimensionality of the signatures; When $d$ is large, successive comparisons will collectively incur an enormous time cost. Although dimension reduction has been extensively studied in the literature (e.g., PCA \cite{Pearson1901}, LSH \cite{Indyk1998}\cite{Gionis1999}, etc.), later experiments show that existing methods significantly degrade the linking accuracy when the dimensionality is highly reduced.

\textit{Curse of cardinality:} 
In practice, the cardinality of the object set, denoted as $n$, is usually massive: reaching millions of candidates. The complexity of the aforementioned pairwise checking method is $O(n \times d)$ which includes lots of unnecessary calculations, since most candidates are actually unpromising in regards to being a member of the $k$-nearest neighbors for the query object. Various algorithms (e.g., \textit{AllPairs} \cite{Bayardo2007}, \textit{APT} \cite{Awekar2009}, \textit{MMJoin} \cite{Lee2010}, \textit{L2AP} \cite{Anastasiu2014}) and indexing structures (e.g., \textit{k-d tree} \cite{Bentley1975KD1}\cite{Robinson1981KDB}\cite{Bentley1979KD2}, \textit{R-tree} \cite{Guttman1984R}\cite{Sellis1987RPlus}\cite{Beckmann1990RStar}, \textit{hB-tree} \cite{Lomet1990HB}\cite{Evangelidis1997HBPi}) have been proposed to scale up the similarity search or $k$-NN search. Nevertheless, later experiments show that directly applying these methods can only achieve limited efficiency improvements.

Our major contributions in this paper include:
%\begin{itemize}
\begin{list}{\labelitemi}{\leftmargin=1em}
	\item An effective way to generate signatures from an object's moving history is proposed. Four representation strategies (i.e., sequential, temporal, spatial, and spatiotemporal) and two quantitative metrics (i.e., commonality and unicity) are studied for their effectiveness in signature construction. High accuracy is achieved for taxi trajectories, which are commonly believed as hard to personalize.
	\item A simple yet effective signature reduction method, CUT, is developed with significantly better performance than the existing algorithms including PCA and LSH based on \textit{spatial shrinking} of moving object footprints.
	\item The high-dimensional object linking problem is transformed to a $k$-NN search in 2D space. A novel indexing structure, WR-tree, is designed to speed up $k$-NN search on signatures. A bulk-loading index construction method is introduced based on a novel optimization criterion of signature enlargement. This index mechanism can support updates (i.e., adding new objects).
	\item Two optimization schemes, re-ranking and stable marriage, are proposed to refine the $k$-NN results and improve the robustness of the linking algorithm.
	%\item We apply two advanced optimization schemes to further polish our linking method. A heuristic re-ranking strategy allows us to quickly narrow the search space of candidates using reduced signatures then conduct finer-grained linking through more informative signatures. Moreover, we apply a flexible adaption of ``stable marriage" which improves the consistency and accuracy of object linking whilst circumventing the intrinsic drawbacks of standard stable marriage methods. Empirical evidence supports the effectiveness of our optimization.
	\item Extensive experiments are conducted on several real-world datasets, demonstrating significantly better accuracy and efficiency by our proposal compared with the state-of-the-art approaches. Object linkability is also shown to be highly sensitive to the signatures, which opens a new approach to trajectory privacy protection.
%\end{itemize}
\end{list}

%The remainder of the paper is organized as follows: Section \ref{sec:problem} introduces some major notations and formally defines the problem of object linking. Section \ref{sec:signature} and Section \ref{sec:linking} describe our proposed strategies for signature construction and object linking, respectively. Extensive experimental results and analysis are reported in Section \ref{sec:experiment}, followed by a brief summary of related work in Section \ref{sec:relatedwork} and a conclusion in Section \ref{sec:conclusion}.
\section{Problem Statement}
\label{sec:problem}
This section introduces some preliminary concepts and formally defines the problem of spatiotemporal entity linking. Table \ref{tab:notation} summarizes several key notations. 

\begin{table}[htp]
	\centering
	\caption{Summarization of key notations.}
	\label{tab:notation}
	\begin{tabular}{|l|p{0.73\columnwidth}|}
		\hline
		\textbf{Notation} & \textbf{Definition} \\
		\hline
		$o$ & a moving object \\
		\hline
		$T(o)$ & the historical trace of a moving object $o$ \\
		\hline
		$f(o)$ & the (reduced) signature of a moving object $o$. $d=|f(o)|$ represents the dimensionality of the signature \\
	
		\hline
		$sim(o_1,o_2)$ & the signature similarity between objects $o_1$ and $o_2$ \\
		\hline
		$k$NN$(q,O)$ & the $k$-nearest neighbors of query object $q$ in the candidate object set $O$, and $n=|O|$ represents the cardinality of the candidate object set \\
		\hline
		$\left\langle g,n(g) \right\rangle$ & $g$ is a $q$-gram and $n(g)$ is the frequency of $g$ in object $o$'s sequential signature $f(o)$ \\
		\hline
		$\left\langle T,n(T) \right\rangle$ & $T$ is a time interval and $n(T)$ is the frequency of $T$ in object $o$'s temporal signature $f(o)$ \\
		\hline
		$\left\langle p,w(p) \right\rangle$ & $p$ is a spatial point and $w(p)$ is the TF-IDF weight of $p$ in object $o$'s spatial signature $f(o)$ \\
		\hline
		$\left\langle h,w(h) \right\rangle$ & $h=(p,T)$ consists of a spatial location $p$ and a time interval $T$, and $w(h)$ is the TF-IDF weight of $h$ in object $o$'s spatiotemporal signature $f(o)$ \\
		\hline                                                                                                   
		$MBR(o)$ & the minimum bounding rectangle over object $o$'s spatial signature $f(o)$ \\
		\hline
	\end{tabular}
\end{table}

\begin{definition}[Spatiotemporal Entity]
	A spatiotemporal entity, denoted as $o$, is a moving object characterized by its position in space that varies over time.
\end{definition}
\begin{definition}[Trace]
	The trace of a spatiotemporal entity (or ``moving object", interchangeably) represents its entire movement history. It is a sequence of spatiotemporal points, denoted as ${T(o)}=\left\langle p_1,p_2,\ldots,p_n\right\rangle$ where each $p=(x,y,t)$ consists of a location $(x,y)$ (e.g., longitude and latitude) at time $t$. Points in a trace are organized chronologically, namely $\forall i<j: p_i.t<p_j.t$.
\end{definition}

The collection of all moving objects is denoted as $O=\{o_1,o_2,\ldots,o_n\}$ with its cardinality $n=|O|$. 
The historical trace of a moving object $o$ can to some extent reflect $o$'s personalized movement pattern, which is valuable for object identification. Logically, two objects with different IDs are possibly the same real-world entity if they share highly similar movement patterns (i.e., \textit{signatures}). 
%Accordingly, we propose the concept of a ``signature'' to facilitate moving object identification.

\begin{definition}[Signature]
	The signature of a moving object $o$, denoted as $f(o)$, is a quantitative representation of $o$'s movement pattern, such that the similarity between two objects $o_1$ and $o_2$ can be measured by the similarity between their corresponding signatures $f(o_1)$ and $f(o_2)$.
\end{definition}

It is assumed that each point in object $o$'s trace $T(o)$ contributes partially to the object's personalized profile: $f(o)$ is constructed from $o$'s entire movement history. Various strategies are proposed to quantify signatures $f(o)$ and calculate signature similarity between two objects $sim(o_1,o_2)$ (Section \ref{sec:signature}). For representation simplicity, $f(o)$ is used hereafter to denote all types of signatures.

Given two sets of moving objects along with their signatures, we aim to identify all pairs of objects that possibly refer to the same real-world entity based on signature similarities. We reduce this task to a $k$-nearest neighbor ($k$-NN) search problem. In other words, for each object in one dataset, a $k$-NN search is conducted in the other dataset to find the matched object.
\begin{definition}[Spatiotemporal Entity Linking]
	Given a query object $q$ and a collection of moving objects $O=\{o_1,$ $o_2,\ldots,o_n\}$, find the top-$k$ nearest neighbors for $q$ with the largest signature similarity $sim(q,o)$.
\end{definition}

% To avoid linking objects with low similarity, the above problem definition can be extended by requiring $sim(q,o)\geq \theta$ for the top-$k$ nearest neighbors of $q$, given a predefined threshold $\theta$.
% In real-world applications, it is possible that the query object cannot be linked to any object in the candidate set due to low similarity. Hence, we can extend the above problem definition by requiring $sim(q,o)\geq \theta$ for the top-$k$ nearest neighbors of $q$, given a predefined threshold $\theta$. 
% This requirement is ignored in the paper, as it does not make much difference to our proposed algorithms.

\section{Signature}
\label{sec:signature}
%An important problem in this work is how to effectively represent and quantify the signature of each moving object to enable accurate and efficient object linking based on signature similarity. 
Inspired by the unique characteristics of trajectories, 
we introduce various methods for signature representation from an object's moving history and propose two heuristic metrics to quantify each signature, as described below:

\textbf{Commonality}. The signature should be representative of an object's movement profile, i.e., ubiquitous throughout the whole trace of the moving object. We define commonality as the frequency of the signature in the object's historical trace.

\textbf{Unicity}. The signature represents an object's personalized movement profile, and should be highly discriminative between candidate objects. We define unicity as the number of unique objects containing the signature, which quantifies the signature's ability to distinguish an object from others.

Given the quantitative signatures of objects, we can estimate their similarities accordingly. We will discuss the technical details in the following sections.

\subsection{Signature Representation}
\label{subsec:representation}
\subsubsection*{1) Sequential Signature} 
Moving objects visit locations in a particular order such that their sequential behavior can become an identifying feature of their historical trajectory. Many existing works on trajectory pattern mining focus on identifying sequential patterns (i.e., a common sequence of locations) from a set of trajectories \cite{Cao2005}\cite{Giannotti2007}, which are then used for real-world applications including routing, location prediction, traffic analysis, etc. This work explores the possibility of utilizing sequential behavior for moving object linking. Specifically, an object $o$'s historical trace $T(o)$ can be regarded as a document, and a set of $q$-grams is extracted from $T(o)$. Each $q$-gram can be associated with its weight in $T(o)$ to identify the most representative and distinctive sequential behaviors of the moving object.

\begin{definition}[Sequential Signature]
  The sequential signature of an object $o$ is a collection of weighted $q$-grams extracted from its historical trace $T(o)$. Specifically, $f(o)=\{\left\langle g_1,w(g_1)\right\rangle, \left\langle g_2,w(g_2)\right\rangle ,\ldots,\left\langle g_d,w(g_d)\right\rangle \}$, where $g$ is a $q$-gram (i.e., a subsequence of $q$ consecutive points), $w(g)$ is its weight in $T(o)$, and $d$ is the total number of distinct $q$-grams in the entire dataset (i.e., \textit{the dimensionality} of $f(o)$) which could be extremely large in practice. We adopt the TF-IDF weighing strategy to calculate $w(g)$, which can capture both commonality and unicity of the sequential signature.
\end{definition}

Given two moving objects $o_1$ and $o_2$ along with their sequential signatures $f(o_1)$ and $f(o_2)$, their similarity is measured using the cosine similarity between the corresponding signatures. To simplify computation, we transform sequential signature into unit length by $L_2$ normalization and then calculate signature similarity by the dot product, i.e., 
\begin{equation}
    \label{eq:sequential_sim}
    \begin{split}
        sim(o_1,o_2)&=\cos(f(o_1),f(o_2))=\frac{f(o_1)\cdot f(o_2)}{||f(o_1)||\times ||f(o_2)||}\\
        &=f(o_1)\cdot f(o_2)=\sum_{i=1}^{d}w_{o_1}(g_i)\times w_{o_2}(g_i)
    \end{split}
\end{equation}
where $w_o(g)$ represents the normalized weight of $q$-gram $g$ in signature $f(o)$. In the rest of the paper, we use $f(o)$ to denote the $L_2$ normalized sequential signature of object $o$.

\subsubsection*{2) Temporal Signature} An individual's moving behavior exhibits some temporal patterns. Consider vehicles: Private cars usually travel at commuter time daily (e.g., 9am and 5pm) whereas taxis might run uninterruptedly all day; Some taxi drivers carry passengers during the daytime whereas others prefer to operate at night. Evidently, temporal travel patterns can help to distinguish moving objects to some extent, motivating the following definition of a temporal signature.

\begin{definition}[Temporal Signature]
    \label{def:temporal_signature}
    The temporal signature of a moving object $o$ is a histogram over equal-size time intervals: $f(o)=[\left\langle T_1,n(T_1)\right\rangle,\left\langle T_2,n(T_2)\right\rangle,\ldots,\left\langle T_d,n(T_d) \right\rangle]$. In particular, we divide one day into $d$ intervals (or bins) where $d$ is the dimensionality of the temporal signature. Each $T_i$ represents a time interval of length $\Delta t=\frac{24}{d}$ hours, and $n(T_i)$ counts the total number of points in $o$'s historical trace $T(o)$ whose timestamp falls into this interval, i.e., $n(T_i)=|\{p|p\in T(o) \land p.t \in T_i\}|$. 
    %Here, $n(T_i)$ also reflects the commonality or representativeness of the signature.
\end{definition}

The similarity between two histograms can be measured by various methods such as correlation, intersection, Chi-square, Battachary distance, KL divergence, $L_p$-norm distance, Earth mover's distance (EMD), etc. We adopt EMD \cite{Tang2013EMD} in this work as it further takes into consideration cross-bin information rather than simply conducting a bin-to-bin matching. 
\begin{comment}
Given two histograms $X=[x_1,\ldots,x_d]$ and $Y=[y_1,\ldots,y_d]$, EMD measures the minimum cost to transform one histogram to the other by transporting elements between bins:
\begin{equation*}
    \begin{split}
    emd(X,Y) & =\min_{F}d(X,Y), \quad s.t., \\
    & \forall i,j\in[1,d]: f_{i,j}\geq 0 \\
    & \forall i\in[1,d]: \sum_{j=1}^d f_{i,j}=x_i \\
    \vspace{-2mm}
    & \forall j\in[1,d]; \sum_{i=1}^d f_{i,j}=y_j
    \end{split}
\end{equation*}
$d(X,Y)$ calculates the total moving cost between $X$ and $Y$ under a flow matrix $F$ and a cost matrix $C$:
\begin{equation*}
    d(X,Y)=\sum_{i=1}^d \sum_{j=1}^d f_{i,j} \times c_{i,j}
\end{equation*}
where $f_{i,j}$ represents the flow (i.e., the amount of elements) to move from $x_i$ to $y_j$, and $c_{i,j}$ indicates the cost of moving the unit flow from the $i$-th bin to the $j$-th bin.
\end{comment}
In order to apply EMD, we transform the temporal signature via $L_1$ normalization (i.e., $n(T_i)=\frac{n(T_i)}{\sum_{j=1}^d n(T_j)}$) as EMD requires the two histograms have the same integral. 
%We consider the $L_1$-normalized temporal signature hereafter, which reflects a probability distribution over time intervals. 
EMD measures the minimum cost ($c_{i,j}$) to transform one histogram to the other by transporting elements between bins. Intuitively, the larger the distance between $i$-th bin and $j$-th bin, the larger the cost $c_{i,j}$. However, the distance between temporal intervals cannot be measured by the number of bins in between. For example, 1:00-2:00 is temporally close to 23:00-24:00 while their bin-wise distance is quite large. Therefore, we design the cost $c_{i,j}$ as below:
\begin{equation*}
    c_{i,j}=\begin{cases}
    \frac{|i-j|\times\Delta t}{12}, & \textit{if }|i-j|\times\Delta t \leq 12\\
    \frac{24-|i-j|\times\Delta t}{12}, & \textit{otherwise}
    \end{cases}
\end{equation*}
%Clearly, $C$ is symmetric and $c_{i,i}=0$. 
Given two objects $o_1$ and $o_2$ with corresponding temporal signatures $f(o_1)$ and $f(o_2)$, we measure their similarity as:
\begin{equation}
    \label{eq:temporal_sim}
    sim(o_1,o_2)=1-emd(f(o_1),f(o_2))
\end{equation}
where $emd(f(o_1),f(o_2))$ is the EMD distance between the two histograms $f(o_1)$ and $f(o_2)$.

\subsubsection*{3) Spatial Signature} In general, a trajectory can be regarded as a special type of time series data whose element represents a geographical location rather than a numeric value. Therefore, besides the sequential and temporal features that exist in general time series, spatial information is also relevant for moving object identification. That is, different moving objects could have different preferences for the locations they visit (residence, restaurant, scenic spot, gas station, etc.). As discovered in \cite{de2013unique}, 4 locations can identify 95\% of the mobile phone users.

\begin{definition}[Spatial Signature]
    \label{def:spatial_signature}
    We define the spatial signature of a moving object $o$ as a weighted vector over the points in its entire historical trace $T(o)$. More specifically, $f(o)=(\left\langle p_1,w(p_1)\right\rangle, \left\langle p_2,w(p_2)\right\rangle,\ldots,\left\langle p_d,w(p_d)\right\rangle)$, where $d$ represents the total number of distinct points in the entire dataset (i.e., the dimensionality of $f(o)$) and $w(p)$ is the weight of point $p$ reflecting its representativeness and distinctiveness. 
\end{definition}

In this work, we adopt the TF-IDF weighing strategy to construct the spatial signature $f(o)$ of moving object $o$ and estimate the importance of each point $p$ in $f(o)$:
\begin{equation*}
\label{eq:signature}
w(p) = tf(p) \times idf(p)
\vspace{-1mm}
\end{equation*}

\begin{itemize}
    \item \textbf{Commonality} measures the frequency of a point in the object's trace, i.e., $tf(p)=\frac{N_{p,T(o)}}{|T(o)|}$ where $N_{p,T(o)}$ is the total number of times $p$ occurs in $T(o)$ and $|T(o)|$ is the total number of points contained in $T(o)$.
    \item \textbf{Unicity} measures how much discriminative information a point provides, i.e., $idf(p)=\log\frac{|O|}{|O_p|}$ where $|O|=n$ represents the total number of moving objects (i.e., the cardinality of candidate object set $O$) and $|O_p|$ represents the number of objects containing $p$ in their traces, namely $O_p=\{o|p\in T(o)\}$.
\end{itemize}

Given two moving objects $o_1$ and $o_2$, their similarity can be measured by the cosine similarity between the corresponding spatial signatures $f(o_1)$ and $f(o_2)$. To simplify the computation, we transform each spatial signature into unit length by $L_2$ normalization and then calculate the signature similarity by the dot product, i.e., 
\begin{equation}
    \label{eq:spatial_sim}
    \begin{split}
        sim(o_1,o_2)&=\cos(f(o_1),f(o_2))=\frac{f(o_1)\cdot f(o_2)}{||f(o_1)||\times ||f(o_2)||}\\
        &=f(o_1)\cdot f(o_2)=\sum_{i=1}^{d}w_{o_1}(p_i)\times w_{o_2}(p_i)
    \end{split}
\end{equation}
where $w_o(p)$ represents the normalized weight of point $p$ in signature $f(o)$. In the rest of the paper, we still use $f(o)$ to denote the $L_2$ normalized spatial signature of object $o$.

\subsubsection*{4) Spatiotemporal Signature} 
Naturally, we can combine multiple features to see if it can achieve better performance. Our empirical results show that sequential signatures are less valuable for object linking. Therefore, we only consider combining spatial and temporal features in this work, i.e., the spatiotemporal signature.

\begin{definition}[Spatiotemporal Signature]
    We define the spatiotemporal signature of a moving object $o$ as a weighted vector over spatiotemporal dimensions $h=(p,T)$, where $p$ represents a spatial location and $T$ is a time interval described in Definition \ref{def:temporal_signature}. We denote the signature as $f(o)=(\left\langle h_1,w(h_1)\right\rangle, \left\langle h_2,w(h_2)\right\rangle,\ldots,\left\langle h_d,w(h_d)\right\rangle)$.
\end{definition}

As in Definition \ref{def:spatial_signature}, we adopt the TF-IDF weighting strategy to calculate $w(h)$. The only difference is that we need to count the total number of times location $p$ occurs in object $o$'s trace $T(o)$ during the time interval of $T$. Combining the two signatures intuitively enlarges the feature space: to alleviate the sparsity of spatiotemporal dimensions, we construct the signature at a coarser granularity via a grid-based spatial representation. That is, $p$ in $h=(p,T)$ represents a grid in the spatiotemporal signature.

All four features (sequential, temporal, spatial, and spatiotemporal) are commonly believed to be very important for various applications of moving object modeling. In this work, we empirically evaluate the four signatures on a real-world taxi dataset.
Interestingly, sequential, temporal, and spatiotemporal signatures are not as effective as the spatial signature for moving object linking.
%In particular, the spatial signature can achieve an average of 82.9\%-85.5\% accuracy for the top-1 nearest neighbor identified from the candidate object set (Detailed results are reported in Section \ref{subsec:effective}). 
Therefore, we will only consider $f(o)$ as the spatial signature hereafter. 
\subsection{Signature Reduction}
\label{subsec:reduction}
%Recall that the signature of a moving object $o$ is constructed from its entire moving history and represented as a TF-IDF weighted $d$-dimensional vector. 
Recall that $f(o)=(\left\langle p_1,w(p_1)\right\rangle, \left\langle p_2,w(p_2)\right\rangle,\ldots,\left\langle p_d,w(p_d)\right\rangle)$. The dimensionality $d$ of the original signature is practically huge considering the large number of points contained in trajectories. This makes it very time-consuming to calculate pairwise signature similarity. Various techniques have been proposed for dimensionality reduction, including principal component analysis (PCA), locality-sensitive hash (LSH), which can be applied to speed up the calculation of cosine similarity and obtain approximately similar signatures. 

\begin{comment}
PCA \cite{Pearson1901} is a classical strategy for dimension reduction, which conducts a linear and orthogonal transformation from a high-dimensional space to a lower dimension. It aims to maximize data variance in the low-dimensional representation by preserving eigenvectors with the largest eigenvalues (i.e., the principal components) from the covariance matrix of the original data. In this way, it can minimize information loss after dimension reduction, and hopefully find most of the top-$k$ nearest neighbors for a query object. 

LSH \cite{Indyk1998}\cite{Gionis1999}, although widely-known for approximate $k$-NN search, can also be regarded as a way of dimension reduction. LSH constructs a family of hash functions $h:R^d\to B$ which map a $d$-dimensional point to a bucket $b\in B$ such that $\forall x,y\in R^d$: 1) if $d(x,y)\leq d_1$, then $P(h(x)=h(y))\geq p_1$ and 2) if $d(x,y)\geq d_2$, then $P(h(x)=h(y))\leq p_2$, where $d_1, d_2, p_1, p_2$ are predefined parameters. We consider \textit{Random Projection} \cite{Andoni2006} or \textit{Super-bit} \cite{Ji2012} (LSH for cosine similarity) in this work. The algorithm transforms a $d$-dimensional weighted vector to a $m$-dimensional bit vector using $k$ random hyperplanes generated from a Gaussian distribution. For each hyperplane, the corresponding hash bit of a vector $x$ is set to 0 or 1 depending on which side of the hyperplane $x$ lies.
\end{comment}

Both PCA \cite{Pearson1901} and LSH \cite{Andoni2006}\cite{Ji2012} are applied in this work for signature reduction in order to speed up the cosine similarity calculation. However, our experimental results in Section \ref{subsubsec:reduction} reveal that we cannot achieve a satisfactory object linking using PCA or LSH if we reduce the signature dimensionality too much (e.g., 10 to 100 dimensions). Fortunately, we observe that signatures exhibit a power-law distribution, meaning that only a small number of points contribute to the majority of the total weights. Therefore, we propose the CUT method (i.e., cutting the long tail) which reduces the original signature by reserving its top-$m$ weighted points only (i.e., updating the weights of all the other points to 0). According to the empirical evaluation in Section \ref{subsubsec:reduction}, such a simple approach to signature reduction can achieve an average accuracy of 80.6\% for taxi linking when the signature dimensionality is reduced from 160,000 to 10 (5\% performance degradation), outperforming PCA and LSH by 79.9\% and 76\% respectively.

The CUT method for dimension reduction naturally results in the side effect of \textit{spatial shrinking} for signatures: the original signature could cover a huge spatial region (vehicles, in particular taxis, could potentially traverse the whole city), while the reduced signature after cutting the long tail is scattered in several small regions (workplace, shopping malls, residence, etc.), since the CUT method only preserves unique locations visited frequently by a moving object. To verify this, we conduct a statistical analysis on a real-world taxi dataset, which compares the pairwise spatial overlap ratio between original signatures with the ratio between reduced signatures. We observe that the overlapping ratio decreases from almost 100\% to 1\% when the original signatures are reduced to 10-dimension. Later sections also discuss the improvement in moving object linking efficiency due to spatial shrinking.

\section{Moving Object Linking}
\label{sec:linking}
Signature reduction helps to eliminate the curse of dimensionality and subsequently speeds up the cosine similarity calculation. 
%For presentation simplicity, we will use ``signature'' and ``reduced signature'' interchangeably in this paper whenever it is clear. 
However, another issue to be considered is the high cardinality of the candidate object set $|O|=n$. Recall that for each object, we need to conduct a $k$-NN search in the candidate set. The naive sequential scan approach is clearly infeasible since $n$ will be extremely large in real applications. In this section, we will first explore the possibility of employing existing search and indexing algorithms to solve the moving object linking problem, and then introduce in detail our proposed indexing structure to further improve the linking efficiency. 

\subsection{Baselines}
\label{subsec:baseline}
\textbf{Baseline 1.} Our problem can be regarded as a cosine similarity search problem. As a result, existing approaches to speeding up cosine similarity search, such as \textit{AllPairs} \cite{Bayardo2007}, \textit{APT} \cite{Awekar2009}, \textit{MMJoin} \cite{Lee2010} and \textit{L2AP} \cite{Anastasiu2014}, can be easily adjusted to solve this problem. Nevertheless, all these algorithms target the curse of dimensionality, i.e., the time cost of calculating cosine similarity based on full-sized vectors. They reduce the number of compared elements at the beginning of the vectors for early candidate pruning. After signature reduction, we suffer more from the high cardinality of the object set rather than the signature size, making these algorithms inappropriate.

\noindent\textbf{Baseline 2.} $k$-NN search has also been extensively studied in Euclidean space, and various spatial indices have been proposed for speeding up $k$-NN search in a relatively high-dimensional Euclidean space, such as \textit{R-tree} \cite{Guttman1984R}\cite{Sellis1987RPlus}\cite{Beckmann1990RStar}\cite{Berchtold1996X}, \textit{k-d tree} \cite{Bentley1975KD1}\cite{Bentley1979KD2}\cite{Robinson1981KDB}, \textit{hB-tree} \cite{Lomet1990HB}\cite{Evangelidis1997HBPi}, etc. Can we transform the $k$-NN search problem under cosine similarity into that in the Euclidean space? The answer is positive based on the following theorem:

\begin{theorem}
	Given two vectors $X=\left\langle x_1,x_2,\ldots,x_d\right\rangle$ and $Y=\left\langle y_1,y_2,\ldots,y_d\right\rangle$ under $L_2$ normalization, their cosine similarity $\cos(X,Y)$ is negatively correlated to their Euclidean distance $euc(X,Y)$.
\end{theorem}

\begin{IEEEproof}
	Note that both $X$ and $Y$ have been $L_2$ normalized to unit length, meaning that $||X||=\sqrt{\sum x_i^2}=1$ and $||Y||=\sqrt{\sum y_i^2}=1$. Therefore, 
	\begin{align*}
		\cos(X,Y)=\frac{X\cdot Y}{||X||\times||Y||}= X\cdot Y=\sum x_iy_i
	\end{align*}
	\vspace{-3mm}
	\begin{align*}
		 euc(X,Y)&=\sqrt{\sum (x_i-y_i)^2}\\
		 &=\sqrt{\sum x_i^2+\sum y_i^2-2\times\sum x_iy_i}\\
		 &=\sqrt{2-2\times\cos(X,Y)}
	\end{align*}
	It can be observed that the larger $\cos(X,Y)$ is, the smaller $euc(X,Y)$ will be. In other words, $\cos(X,Y)$ is negatively correlated to $euc(X,Y)$.
\end{IEEEproof}

To address the object linking problem, we regard each object's signature $f(o)$ as a $d$-dimensional point and utilize the existing spatial indices to find the $k$-nearest neighbors in $O$ that minimize the Euclidean distance $euc(f(q),f(o))$ as the linking candidates for the query object $q$. However, previous work has revealed that some spatial indices (e.g., R-tree and k-d tree) are practically efficient only when the dimensionality $d$ is not too large (e.g., $d\leq 20$). In a sufficiently high-dimensional situation (e.g., $d>20$), approximate algorithms, such as LSH \cite{Indyk1998}\cite{Gionis1999}, are widely-adopted alternatives which can obtain near-optimal results for $k$-NN search with a high probability.

\subsection{Weighted R-tree}
In the above baselines, we regard a signature purely as a weighted vector while ignoring the information associated with each element. In fact, the signature is composed of a collection of spatial points in the 2D space. 
%Is it possible to utilise this information to further speed up moving object linking? 
In this section, we present in detail our novel indexing structure which we call \textit{Weighted R-tree (WR-tree)}, and show that better efficiency can be achieved by taking both spatial and weight information into consideration.

\subsubsection{Pruning Strategies}
We first introduce two pruning strategies which can help to rule out unpromising candidates earlier.

\subsubsection*{1) Pruning by Spatial Overlapping} 
Our indexing scheme is based on the locality assumption that each moving object usually travels in a limited region. For example, individuals with a nine-to-five working pattern are always subject to routes between residences and workplaces during weekdays. Taxi drivers, although traveling according to customers' requirements, have the flexibility to choose preferred riding regions. 
%Someone would like to travel around the downtown of a city or drive to and from the railway stations. 
In other words, we can bound an object $o$'s signature, in particular spatial points in the signature, with a minimum bounding rectangle (MBR) denoted as $MBR(o)$. We say $p \in MBR(o)$ if point $p$ is spatially covered by $MBR(o)$, and $MBR(o_1) \cap MBR(o_2)\neq\emptyset$ if two MBRs $MBR(o_1)$ and $MBR(o_2)$ spatially overlap with each other. Note that $MBR(o_1) \cap MBR(o_2)\neq\emptyset$ only specifies the spatial overlapping between two MBRs, which does not necessarily mean the corresponding signatures intersect with each other.
\begin{theorem}
	Given two moving objects $o_1$ and $o_2$ and their minimum bounding rectangles $MBR(o_1)$ and $MBR(o_2)$, if $MBR(o_1) \cap MBR(o_2)=\emptyset$, then $sim(o_1,o_2)=0$.
\end{theorem}

\begin{comment}
\begin{IEEEproof}
	We prove the theorem by contradiction. According to Eq. \ref{eq:spatial_sim}, $sim(o_1,o_2)=\sum_{i=1}^{d}w_{o_1}(p_i)\times w_{o_2}(p_i)$, which is an accumulation of weight production for points whose weights in both $f(o_1)$ and $f(o_2)$ are larger than 0. Assume that $sim(o_1,o_2)>0$. This means there exists at least one point $p$ such that $w_{o_1}(p)>0 \land w_{o_2}(p)>0$. Based on our signature reduction strategy, $w_o(p)>0$ if and only if point $p$ is reserved in $o$'s reduced signature or in other words, $p$ is covered by $MBR(o)$, i.e., $p\in MBR(o)$. It follows directly that $sim(o_1,o_2)>0$ means there exists at least one point $p$ such that $p\in MBR(o_1) \land p \in MBR(o_2)$, which contradicts with the condition that $MBR(o_1) \cap MBR(o_2)=\emptyset$.
\end{IEEEproof}
\end{comment}

Accordingly, we can skip calculating the cosine similarity between query $q$ and the candidate object $o$ if their MBRs do not overlap with each other. Checking for MBR overlap is intuitively faster in comparison to calculating cosine similarity. Our statistical analysis about the spatial shrinking effect of CUT signature reduction (i.e., only 1\% overlapping ratio between reduced signatures) suggests that the spatial pruning strategy could be very effective for speeding up moving object linking.

\noindent\textbf{Baseline 3}. 
A straightforward way of utilizing this pruning strategy is to calculate the MBRs of all the candidate objects using the R-tree indexing scheme. Given a query MBR, we conduct a \textit{range query} on the R-tree to find all the candidate MBRs overlapping with the query MBR. We then calculate the cosine similarities between these candidates and the query object one by one, and determine $k$-nearest neighbors based on the ranking of cosine similarities. This approach can speed up $k$-NN search as it eliminates many unnecessary similarity calculations. Nevertheless, each point in the object signature is associated with not only the spatial information but also the weight. Is it possible to combine both types of information to improve search efficiency?

\subsubsection*{2) Pruning by Signature Similarity}
We introduce a method to further prune the $k$-NN search by considering the signature similarity. Assume that multiple signatures, $f(o_1),f(o_2),\ldots,f(o_m)$, are aggregated into one signature denoted as $f(o_1,o_2,\ldots,o_m)$ such that:
\begin{itemize}
	\item Point set of the aggregated signature is a union of point sets of the constituting signatures, i.e., $p \in f(o_1,o_2,\ldots$, $o_m)$ iff $\exists o\in\{o_1,o_2,\ldots,o_m\}, p \in f(o)$;
	\item Each point weight in the aggregated signature is the maximum value of the corresponding point weights from the constituting signatures, i.e., $w(p)=\max\{w_{o_1}(p)$, $w_{o_2}(p),\ldots,w_{o_m}(p)\}$.
\end{itemize}
\begin{theorem}
	Given an aggregated signature $f(o_1,\ldots,o_m)$ and a query signature $f(q)$, if $f(o_1,\ldots,o_m)\cdot f(q)<\theta$, then $\forall o\in\{o_1,o_2,\ldots,o_m\}, sim(o,q)<\theta$.
\end{theorem}
\begin{IEEEproof}
	Consider any object $o\in\{o_1,o_2,\ldots,o_m\}$. Based on the signature similarity defined in Eq. \ref{eq:spatial_sim},
	\begin{align*}
		sim(o,q)&=\sum w_o(p)\times w_q(p)\\
		&\leq \sum \max\{w_{o_1}(p),w_{o_2}(p),\ldots,w_{o_m}(p)\}\times w_q(p)\\
		&=f(o_1,o_2,\ldots,o_m)\cdot f(q)<\theta
	\end{align*}
	This ends the proof of Theorem 3.
\end{IEEEproof}

Therefore, we can ignore checking objects $o_1,o_2,\ldots,o_m$ if their aggregated signature is not a promising $k$-NN candidate for the query object $q$.

\subsubsection{WR-tree structure and $k$-NN search}
The basic idea of our indexing scheme is to cluster the set of candidate objects into several disjoint subsets and aggregate them into multiple hierarchies, so that we can prune unpromising subsets earlier based on both spatial overlapping and signature similarity. To this end, we propose the Weighted R-tree index, which incorporates point weights into the R-tree structure. 

\begin{figure}[htp]
	\centering
	\includegraphics[width=0.48\textwidth]{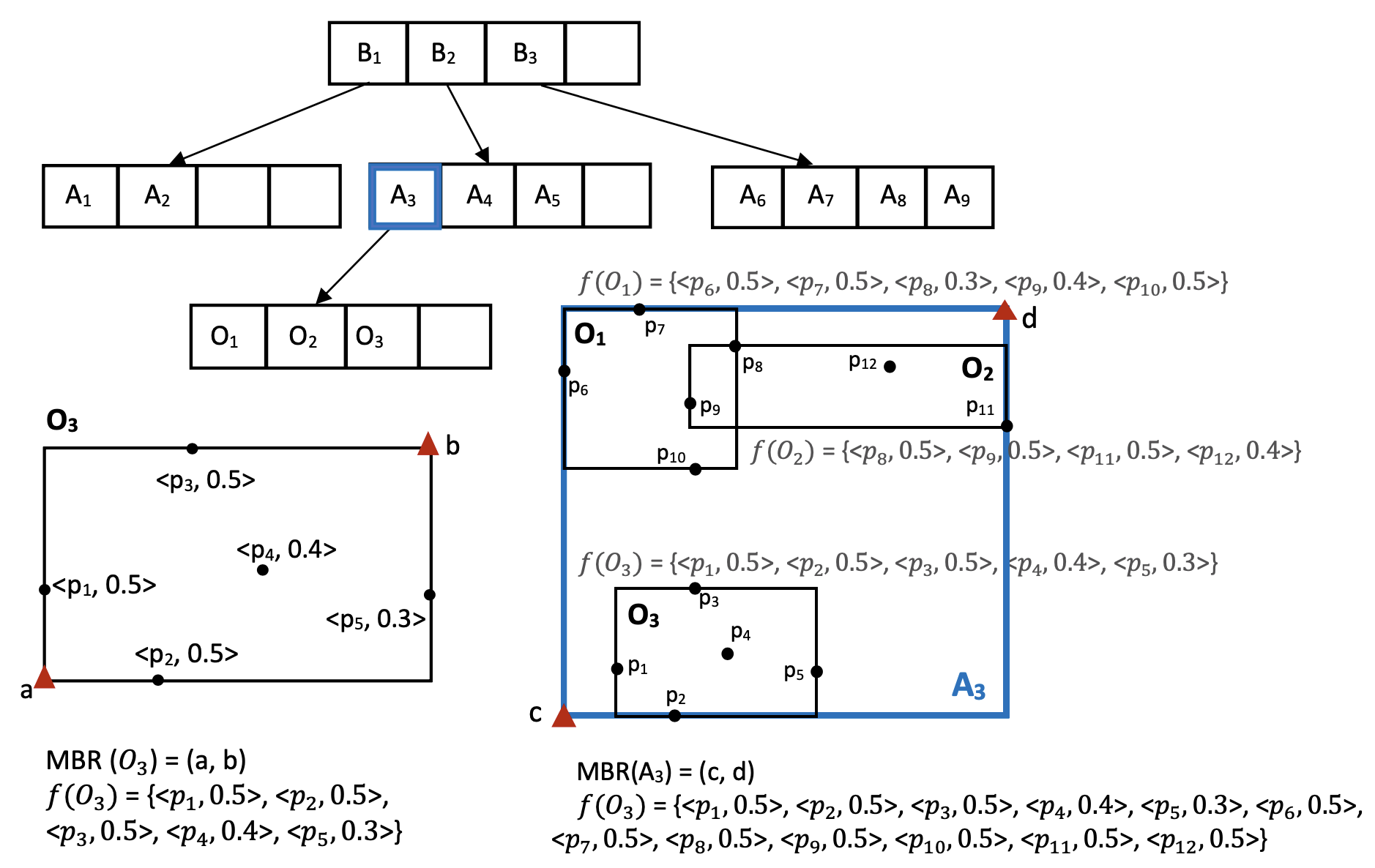}
	\caption{An example of the WR-tree.}
	\label{fig:wrtree}
\end{figure}

Fig. \ref{fig:wrtree} illustrates an example of the WR-tree. Each MBR at the leaf describes a single moving object, while MBRs at higher levels represent the aggregation of all the child nodes. 
Leaf nodes store three fields: 1) object identifier, 2) object signature, and 3) object MBR. MBRs are represented using two points, e.g., $MBR(o_3)=(a,b)$. Intermediate nodes also store three fields: 1) pointers to child nodes, 2) aggregated signature of child nodes, and 3) MBR corresponding to the aggregated signature. For example, since $p_8$ occurs in both $f(o_1)$ and $f(o_2)$ with different weights $0.3$ and $0.5$ respectively, the weight of $p_{8}$ in $f(A_3)$ takes the largest value $0.5$.

\begin{algorithm}
	\caption{$k$-NN search on WR-tree}
	\label{algo:wrsearch}
	\scriptsize
	\KwIn{$T$: WR-tree, $q$: query object}
	\KwOut{$kNN$: $k$-nearest neighbors}
	$kNN=\emptyset$, $kSim=0$, $Q=\emptyset$; \\
	$Q.push(\left\langle root(T),d\right\rangle )$; \\
	\While{$Q\neq\emptyset$}{
		$\left\langle u,sim(u,q)\right\rangle =pop(Q)$; \\
		\If{$sim(u,q)\leq kSim$}{
			break;
		}
		$V=child(u)$; \\
		\If{$V=\emptyset$}{
			$kNN.insert(\left\langle u,sim(u,q)\right\rangle )$; \\
			$kSim=kNN[k].sim$;
		}
		\Else{
			\For{$v\in V$}{
				\If{$MBR(v)\cap MBR(q)\neq\emptyset$}{
					$sim(v,q)= f(v)\cdot f(q)$; \\
					\If{$sim(v,q)>kSim$}{
						$Q.push(\left\langle v,sim(v,q)\right\rangle )$;
					}
				}
			}
		}
	}
	\Return{$kNN$};	
\end{algorithm}

Algorithm \ref{algo:wrsearch} describes the overall process of $k$-NN search on the WR-tree considering both pruning methods. Given a query object $q$ along with its signature and MBR, we search for $k$-nearest neighbors in the WR-tree using a ``best-first'' strategy. In particular, we use a priority queue $Q$ to arrange candidate tree nodes in descending order of their signature similarities with $q$. Note that the signature of a non-leaf node is the aggregated signature whose similarity with $f(q)$ could be larger than 1. The priority queue is initialized with the root node and similarity $d$ (line 2. $d$ is the dimensionality of a signature vector, and hence the upper bound of signature similarity). In each iteration, we pop the top tree node $u$ from $Q$ which has the currently largest $sim(u,q)$ (line 3) and check whether $u$ is a leaf node or an intermediate node. If $u$ is a leaf node, a moving object has been retrieved. We insert $u$ into the list of $k$-nearest neighbors $NN$ if $sim(u,q)>kSim,$ where $kSim$ denotes the signature similarity of the $k$-th neighbor with $q$, and then update $kSim$ (lines 8-10). If $u$ is an intermediate node, we retrieve all its child nodes and add promising children into the queue for further checking (lines 11-16). During this process, the aforementioned rules are used to identify unpromising child node $v$ and prune the corresponding search branch:
\begin{itemize}
	\item Pruning by spatial overlapping: If $MBR(v)\cap MBR(q)=\emptyset$,  this branch can be pruned without calculating the signature similarity (line 13);
	\item Pruning by signature similarity: If $sim(v,q)\leq kSim$, all its child nodes are unpromising for $k$-NN search and can be pruned (line 15).
\end{itemize}
The $k$-NN search continues until the priority queue $Q$ is empty (line 3) or the top tree node in $Q$ is unpromising (lines 4-6, all the remaining nodes in $Q$ and their children are unlikely to achieve a similarity larger than $kSim$).

\subsubsection{WR-tree Construction and Update}
\label{subsec:update}
An important factor that needs to be carefully considered when constructing WR-tree is how to hierarchically cluster candidate objects to optimize pruning power. In traditional R-trees, minimization of both \textit{coverage} (i.e., MBRs do not cover too much empty space) and \textit{overlap} (i.e., MBRs do not share too much common space so that fewer subtrees need to be processed during search) is vital to the performance. 

In the WR-tree, however, we should also consider the size of the aggregated signature due to the following reasons: 
1) The aggregated signature is stored in every internal node, which means we can reduce the storage cost of WR-tree if the size of each aggregated signature is minimized; 
2) An upper bound needs to be calculated when pruning an internal node and the corresponding subtree by signature similarity. The upper bound is defined as the cosine similarity between the query object's signature and the aggregated signature stored in the internal node. Therefore, the size of the aggregated signature will affect the computational cost of $k$-NN search. 

\begin{figure}[htp]
\centering
\includegraphics[width=0.3\textwidth]{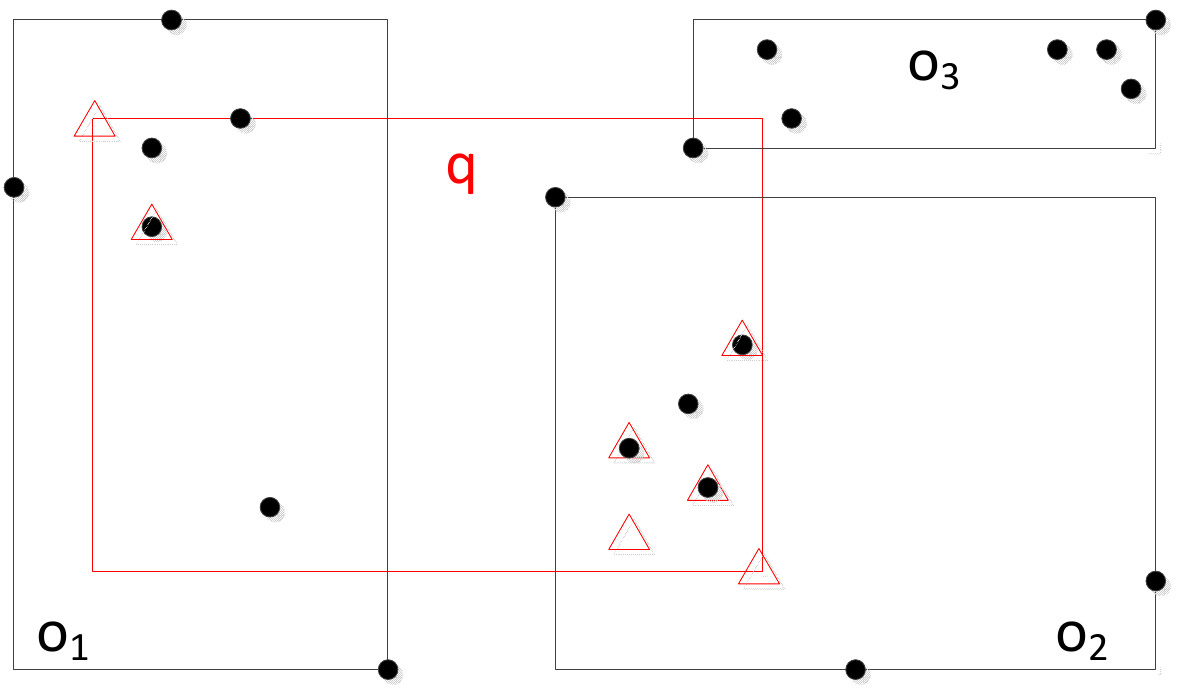}
\caption{An example of signature enlargement. Rectangles represent MBRs, dots and triangles represent points contained in the signature. A dot and a triangle correspond to the same point when overlapping.}
\label{fig:wrtree_construction}
\end{figure}

When constructing the WR-tree, we consider the following heuristic criteria for optimization:
\begin{itemize}
	\item Minimize the size of the aggregated signature.
	\item Minimize the area covered by the MBR.
	\item Minimize the overlapping between MBRs.
\end{itemize}
In other words, we further examine the \textit{signature enlargement} when merging two nodes, in addition to area enlargement and overlapping enlargement. A natural way of reducing signature enlargement is to maximize the number of common points between two signatures. As illustrated in Fig. \ref{fig:wrtree_construction}, the signature $f(q)$ spatially overlaps with all the three signatures $f(o_1)$, $f(o_2)$ and $f(o_3)$. Although combining $f(q)$ and $f(o_1)$ results in a smaller area enlargement than merging $f(q)$ and $f(o_2)$, the signature enlargement increases on the contrary since $f(q)$ shares less common points with $f(o_1)$ than $f(o_2)$.

We propose a bulk-loading algorithm for building WR-tree based on the strategy known as Sort-Tile-Recursive (STR) \cite{leutenegger1997str}. It builds the WR-tree in a bottom-up manner where tree nodes are recursively merged into upper levels. The total number of leaf nodes is initialized as $l=\lceil\frac{n}{c}\rceil$, where $n$ is the number of objects and $c$ is the node capacity. We spatially sort the objects based on the longitude (resp. latitude) and divide them into several equal-size partitions with the size of $s=\lceil \frac{n}{l} \rceil$ (resp. $s=\lceil \frac{n}{\sqrt{l}} \rceil$). Objects in each partition are merged into tree nodes. If the bulk size $s$ exceeds the node capacity $c$, they will be recursively bulk-loaded using the same method. Algorithm \ref{algo:build} depicts our implementation of the \textit{MergeNode()} function. Given a bulk of spatially ordered objects, each tree node is initialized by the first unassigned object (line 10) and the remaining $c-1$ objects are selected from the bulk by minimizing signature enlargement (i.e., maximizing the number of common points, lines 11-13). 
Compared with the insertion-based index construction adopted in \cite{jin2019moving}, this bulk-loading based WR-tree performs significantly better, as demonstrated in our experimental results in Section \ref{subsubsec:index}.

\begin{algorithm}[hbt]
	\caption{MergeNode()}
	\label{algo:build}
	\scriptsize
	\KwIn{$L$: the list of spatially ordered objects in a bulk; $c$: node capacity}
	\KwOut{$N$: newly created nodes}
	$N = \emptyset$;\\
	\If{$|L| < c$}{
	    $node = createNode(L);$\\
	    $N.add(node);$\\
	}
	\Else{
	    %$c = \lceil c \cdot \gamma \rceil;$ \\
    	$n = \lceil |L|/c \rceil;$  \textit{// \# nodes to be created} \\
	    %$c = \lceil|L|/n \rceil;$ \textit{// \# elements to be assigned to each single node} \\
	    $idx = 0$; \\
	    \While{$|N|<n$}{
	        $node = createNode();$\\
	        $node.addElement(L.get(idx));$\\
	        \While{$node.size() < c$}{
	            $e = MaxCommonPoints(node, L);$\\
	            $node.addElement(e);$\\
	        }
	        $N.add(node);$\\
	        $Update(idx, L);$  \textit{// get the index of next available element in $L$} \\
	    }
	}
	\Return{$N$};\\
\end{algorithm}

\vspace{-1mm}
\begin{algorithm}[hbt]
	\caption{UpdateSubtree()}
	\label{algo:choose}
	\scriptsize
	\KwIn{$node$: root of the subtree to be updated; $o$: object to be inserted}
	%\KwOut{$target$: the leaf node where $o$ will be inserted}
	$Update(node, o)$;\\
	$C = child(node);$\\
	$N = MaxCommonPoints(o, C);$\\
	\If{$|N| >1$}{
		$N = MinArea(o, N)$;\\
		\If{$|N| >1$}{
			$N = MinOverlap(o, N)$;\\
		}
	}
	Randomly select a node from $N$ and regard it as $target$;\\
	\textit{// Tree nodes are updated recursively until the leaf}\\
	\If{$target.level = 1$}{
		$Update(target, o)$;\\
		%\Return {$target$};\\
	}
	\Else{
		%\Return{ChooseSubtree($target$, $o$)};\\
		$UpdateSubtree(target, o)$;\\
	}
\end{algorithm}

A WR-tree is useful in practice only if it can append new objects after construction while maintaining an acceptable arrangement of the tree nodes to maximize pruning power. We adopt an incremental strategy to update the WR-tree. Specifically, when inserting a new object, we traverse down to the leaf node by recursively choosing the most attractive subtree based on the three heuristic criteria discussed above. Algorithm \ref{algo:choose} describes our implementation of the \textit{UpdateSubtree()} function, where we select the subtree by maximizing the number of common points between signatures and then minimizing the area and overlapping between MBRs. Tree nodes along the traversal path are updated with respect to the corresponding aggregated signature and MBR. However, we observe that the WR-tree gradually becomes less efficient with more and more new objects being inserted beyond the initial bulk-loading. According to our empirical results in Section \ref{subsubsec:index}, querying a WR-tree that has been updated using incremental insertions for several times is less efficient compared to a newly bulk-loaded tree indexing the same objects. Therefore, a more practical solution would be to periodically rebuild the WR-tree using bulk-loading so that objects are arranged globally and evenly in the tree.

\section{Robustness and Linkability}
\label{sec:discussion}
In this section, we introduce two optimization methods to further improve the accuracy and robustness of object linking. We also discuss the possibility of using our designed signature for trajectory privacy protection.

\subsection{Improving Linking Accuracy}
\label{subsec:optimization}
The major motivation of signature reduction is to mitigate the curse of dimensionality and thus improve the linking efficiency. However, a drawback of using the reduced signatures is an observed drop in the linking accuracy of the $k$-NN search results. This is because the reduced signatures essentially encode less information than larger signatures, making it more difficult to link objects. We introduce two optimization schemes to mitigate this drawback without significantly impacting the efficiency. Note that these optimization schemes are generalized and can be applied to any of the aforementioned linking approaches.

\subsubsection*{1) Re-Ranking Strategy}
Naturally, the matched objects are much more likely to be retrieved by a top-$k$ ($k>1$) NN search than a stricter top-1 NN search. In other words, with the limited information encoded in a reduced signature, the correct linking might appear at a lower rank than the top-1 result, meaning that a further refinement of the $k$-NN search results is needed. To this end, we introduce a heuristic \textit{re-ranking} strategy to improve the accuracy of stricter linking queries. Specifically, given a query object $q$, we first retrieve the top-$k$ ($k>1$) candidate objects using signatures of small size (e.g., $d$=10), and then re-order those objects by calculating their signature similarities using larger and more informative signatures (e.g., $d$=500). It is obvious that the pre-filtering phase is quite efficient by $k$-NN search with small-size signatures, and the extra computational cost is also acceptable by limiting the re-ranking only in the top-$k$ candidates.

%Empirical observations illustrate that objects are much more likely to be retrieved by a top-5 NN search than a stricter top-1 NN search. In other words, with the limited information encoded by a reduced signature, candidate objects that are not the correct object can sometimes have a higher similarity to the query object. To this end, we adopt a heuristic \textit{re-ranking} strategy to improve the accuracy of stricter linking queries. Specifically, we first narrow the search space by performing a looser search, e.g., top-5 NN using reduced signatures, as usual. Afterward, we re-rank the retrieved top-5 neighbors by comparing their more informative full-sized signatures. Based on the experimental results shown in Table \ref{tab:optimisation_acc} and \ref{tab:optimisation_time}, applying such a two-stage approach greatly improves the linking accuracy when only considering the top candidate (i.e., $Acc@1$) with acceptable additional computation costs.

\subsubsection*{2) Stable Marriage Algorithm}
Our second optimization scheme extends the classic \textit{stable marriage} problem (SMP) \cite{Gale1962}. Given two equal-size datasets, stable marriage aims to find a one-to-one mapping between the elements of each dataset. Elements are mapped based on a given order of preferences for elements in their opposing dataset, and the final mapping of elements must be stable, i.e., no two unpaired elements would rather be paired with each other. 

In practice, a guaranteed stable marriage requires a complete pairwise computation of the preference lists for every element, incurring an unacceptable time cost. Additionally, the problem requires two equally sized datasets, which may not necessarily be the case. Hence, traditional algorithms for stable marriage cannot be directly applied to our object linking problem, even though signature similarity can naturally be used to rank preferences. In this work, we modify the stable marriage algorithm to explore the most similar candidate for each object and guarantee a relatively stable matching between two sets of moving objects. As shown in Algorithm \ref{algo:marriage}, our stable marriage method is based on each object's top-$k$ NNs, which have been quickly generated by using any of the aforementioned linking algorithms on the reduced signatures. For each object, its set of top-$k$ NNs along with their signature similarities form a preference list which will be used in the following \textit{proposal} phase (line 8). The process of \textit{proposal} and \textit{engage} repeats until no stable match can be obtained further (lines 3-20). Finally, all the unmatched objects will use the original $k$-NN results as their linking (lines 21-23). Our empirical results in Section \ref{subsubsec:accuracy} verify that stable marriage can effectively improve accuracy of the top-ranked object linking, while incurring limited extra computational cost.
%Moreover, for our implementation, the sizes of the two datasets do not have to be equal, to be more applicable in practice.

\begin{algorithm}[tb]
	\caption{$k$-NN Based Stable Marriage}
	\label{algo:marriage}
	\scriptsize
	\KwIn{$O, O^\prime$: two sets of objects to be linked together; $kNN, kNN^\prime$: two sets of top-$k$ NNs for each object from both sets;}
	\KwOut{$P$: the stable matching pairs between two sets}
	$P=\emptyset$;\\
	$Q=\emptyset$; \textit{// the set of unmatched objects}\\
	\While{$O \neq \emptyset$}{
		$o_i = O.poll()$;\\
		$o_m^\prime = kNN_i.poll()$;\\
		\If{$P.contains(o_m^\prime)$}{
			$o_j=P.get(o_m^\prime)$;\\
			\If{$getRank(o_i, kNN_m^\prime) > getRank(o_j, kNN_m^\prime)$}{
			%\If{$sim(o_i, o^\prime) > sim(o_j,o^\prime)$}{
				$P.add(o_i, o_m^\prime)$;\\
				\If{$kNN_j \neq \emptyset$} {
					$O.add(o_j)$;\\
				}
				\Else{
					$Q.add(o_j)$;\\
				}
			}
			\Else{
				\If{$kNN_i \neq \emptyset$}{
					$O.add(o_i)$;\\
				}
				\Else{
					$Q.add(o_i)$;\\
				}
			}
		}
		\Else{
			$P.add(o_i, o_m^\prime)$;\\
		}
	}
	\textit{// Handle the remaining objects which fail to be matched before}\\
	\For{$o_i \in Q$}{
		$P.add(o_i, getTopOne(kNN_i))$;\\
	}
	\Return $P$;\\
\end{algorithm}

\subsection{Trajectory Linkability}
\label{subsec:privacy}
Trajectories can disclose highly-sensitive information of an individual, such as personal gazetteers and social relationships. Even when trajectories are anonymized (namely, IDs removed), the possibility of object re-identification through spatiotemporal entity linking, as studied in this work, still exposes a high risk of privacy leaks. Although much effort has been devoted to adapting the existing privacy protection models to trajectory dataset, such as $k$-anonymity \cite{abul2010anonymization}\cite{gramaglia2015hiding}, $l$-diversity and $t$-closeness \cite{tu2018protecting}, differential privacy \cite{he2015dpt}, and plausible deniability \cite{bindschaedler2016synthesizing}, they either incur massive computational cost \cite{gramaglia2015hiding} or cannot achieve a satisfactory protection against the linking attack. 

In this work, we first explore the extent of linkability, especially how much data is needed to achieve a reliable linking. Our empirical result in Figure \ref{fig:accuracy_days} shows that signatures derived from about one-week trajectories are enough to identify the majority of individuals ($>75\%$). Moreover, we study the possibility of addressing this linkability attack using our proposed signature design. We introduce a solution called \textit{signature closure}, which iteratively suppresses the reduced signature (i.e., the top-$m$ TF-IDF weighted points) of a moving object from its historical trace and releases the modified trajectories. Our experimental results in Section \ref{subsubsec:linkability} demonstrate that objects with modified traces cannot be easily re-identified. It is worth noting that such modification will not affect much of the utility as evidenced by Table \ref{tab:closure}, since only a very limited amount of information (i.e., a small set of points) is removed from the original trajectories.
\section{Experiments}
\label{sec:experiment}
We conduct extensive experiments on a real-world dataset to evaluate the performance of our algorithms. We report the experimental results and analysis in this section.

\subsection{Experiment Setting}
\label{subsec:setting}
\textbf{Dataset.} 
We use a real-life dataset of 359,666,430 GPS points generated by 13,132 taxis in Beijing city over one month. In order for the signature to capture enough information about a taxi's moving pattern, we remove taxis with less than 7,000 points and finally reserve 12,000 taxis. Each taxi is associated with a trace which is a concatenation of all its trajectories in chronological order. We use a commercial map of Beijing to extract road intersections and align traces by trajectory calibration \cite{Su2013}. 181,265 intersections are discovered, which is regarded as the dimensionality of the original signature.
%We regard road intersections as anchor points for trajectory calibration since they can well preserve the shape or direction change of a trajectory. To this end, we use a commercial map of Beijing city as the reference and extract 296,709 road intersections. 181,265 road intersections are remained after trajectory calibration (i.e., these intersections have at least one taxi passing by), which is regarded as the dimensionality of the original signature. 
We then randomly sample 3000, 6000, 9000 taxis respectively, from the original dataset, to evaluate the performance of moving object linking on different sizes of datasets. We also evaluate our proposals on another publicly available dataset, Geolife \cite{zheng2011geolife}, which was released by Microsoft Research Asia and has been widely utilized in various works on user linking and trajectory privacy protection. Geolife contains check-ins generated by 178 users of the Geolife social networking service in a period of over four years (from April 2007 to October 2011). After pre-processing, 175 users are remained in our test dataset with 20,828,028 GPS points in total. Since the results on Geolife share a quite similar trend with those obtained on the Beijing taxi dataset, we only report the performance of the signature representation for user linking and the trajectory privacy protection on the Geolife dataset in this paper.

\noindent\textbf{Evaluation metrics.}
We divide the original dataset into two parts $Q$ and $D$. Each object exists in both datasets and its trace consists of only half of the original trajectory set. We assign trajectories to $Q$ and $D$ alternately to eliminate the influence of temporal dynamics (e.g., $Q$ and $D$ include all trajectories in odd days and even days respectively). For each object $q$ in dataset $Q$, we conduct a $k$-NN search in dataset $D$ and check if the same object appears in the top-$k$ neighbors. We run this operation $|Q|$ times, and report the average accuracy.

\begin{equation}
\label{eq:accuracy}
Acc@k = \frac{|Q^*_k|}{|Q|}
\end{equation} 
In Eq. \ref{eq:accuracy}, $Q^*_k$ represents the set of query objects that can successfully find themselves in the corresponding top-$k$ results, namely, $Q^*_k=\{q|q\in kNN(q,D)\}$.

\noindent\textbf{Algorithms.} Our performance evaluation has two parts: accuracy and efficiency.
%has two parts: 1) effectiveness of the signature representation and signature reduction, and 2) efficiency of moving object linking algorithms. 
For accuracy evaluation, we compare the four signature representations introduced in Section \ref{subsec:representation}: sequential, temporal, spatial, and spatiotemporal. We then report $Acc@k$ achieved by the original spatial signature and compare it with the signature reduction strategies discussed in Section \ref{subsec:reduction}: CUT, PCA, and LSH (We control the reduced dimensionality $m$ of LSH by the number of hash functions). We also evaluate the effectiveness of the optimization methods introduced in Section \ref{subsec:optimization}: re-ranking, stable marriage, and their combined effect. For the efficiency evaluation, we compare WR-tree with all baselines discussed in Section \ref{sec:linking}, and report the performance of index build and update.
\begin{comment}
\begin{itemize}
    \item Linear: $k$-NN search by a linear scan in dataset $D$.
    \item Baseline 1: Efficient cosine similarity search algorithms (e.g., \textit{L2AP} \cite{Anastasiu2014}). We determine threshold $\theta$ based on the average similarity of $k$-th neighbors for all objects in $Q$.
    \item Baseline 2: $d$-dimensional $k$-NN search with LSH. Since LSH only supports approximate $k$-NN search, we report the time cost of LSH when it achieves similar linking accuracy with the other algorithms.
    \item Baseline 3: $2D$ range search with R-tree, followed by a one-by-one similarity calculation.
\end{itemize}
\end{comment}

All the above algorithms are implemented in Java, and all the experiments are conducted on a server with two Intel(R) Xeon(R) CPU E5-2630, 10 cores/20 threads at 2.2GHz each, 378GB memory, and Ubuntu 16.04 operating system.

\begin{table*}[htb]
	\centering
	\scriptsize
	\renewcommand\arraystretch{1.2}
	\caption{Effectiveness of signature representation ($|D|=3000$).}
	\label{tab:accuracy_representation}
	\begin{tabular}{|p{1.1cm}<{\centering}|p{0.6cm}<{\centering}|p{0.6cm}<{\centering}|p{0.6cm}<{\centering}|p{0.6cm}<{\centering}|p{0.6cm}<{\centering}|p{0.6cm}<{\centering}|p{0.6cm}<{\centering}|p{0.6cm}<{\centering}|p{0.6cm}<{\centering}|p{0.6cm}<{\centering}|p{0.6cm}<{\centering}|p{0.6cm}<{\centering}|p{0.65cm}<{\centering}|p{0.7cm}<{\centering}|p{0.7cm}<{\centering}|p{0.7cm}<{\centering}|}
		\hline
		Methods & \multicolumn{5}{c|}{Sequential ($q$)} & \multicolumn{7}{c|} {Temporal ($\Delta t$)} & Spatial & \multicolumn{3}{c|}{Spatiotemporal (\# of grids)} \\
		\hline
		Parameters & 1 & 2 & 3 & 4 & 5 & 1h & 2h & 3h & 4h & 6h & 8h & 12h & N/A & $100^2$ & $200^2$ & $300^2$\\
		\hline
		$Acc@1$ & 0.855 & 0.804 & 0.773 & 0.737 & 0.704 & 0.127 & 0.123 & 0.104 & 0.087 & 0.042 & 0.018 & 0.004 & 0.855 & 0.535 & 0.567 & 0.583  \\
		\hline
		$Acc@2$ & 0.904 & 0.862 & 0.838 & 0.811 & 0.791 & 0.169 & 0.167 & 0.145 & 0.124 & 0.074 & 0.033 & 0.007 & 0.904 & 0.587 & 0.613 & 0.630 \\
		\hline
		$Acc@3$ & 0.928 & 0.892 & 0.871 & 0.848 & 0.829 & 0.195 & 0.186 & 0.172 & 0.150 & 0.092 & 0.046 & 0.009 & 0.928 & 0.612 & 0.640 & 0.651 \\
		\hline
		$Acc@4$ & 0.940 & 0.913 & 0.891 & 0.872 & 0.854 & 0.216 & 0.205 & 0.198 & 0.174 & 0.113 & 0.057 & 0.011 & 0.940 & 0.632 & 0.659 & 0.681 \\
		\hline
		$Acc@5$ & 0.948 & 0.924 & 0.905 & 0.887 & 0.869 & 0.233 & 0.220 & 0.216 & 0.192 & 0.131 & 0.071 & 0.013 & 0.948 & 0.647 & 0.673 & 0.693 \\
		\hline
	\end{tabular}
\end{table*}

\begin{table*}[htb]
	\centering
	\scriptsize
	\renewcommand\arraystretch{1.2}
	\caption{Effectiveness of signature representation on Geolife ($|D|=175$). }
	\label{tab:accuracy_geolife}
	\begin{tabular}{|p{1.1cm}<{\centering}|p{0.6cm}<{\centering}|p{0.6cm}<{\centering}|p{0.6cm}<{\centering}|p{0.6cm}<{\centering}|p{0.6cm}<{\centering}|p{0.6cm}<{\centering}|p{0.6cm}<{\centering}|p{0.6cm}<{\centering}|p{0.6cm}<{\centering}|p{0.6cm}<{\centering}|p{0.6cm}<{\centering}|p{0.6cm}<{\centering}|p{0.65cm}<{\centering}|p{0.7cm}<{\centering}|p{0.7cm}<{\centering}|p{0.7cm}<{\centering}|}
		\hline
		Methods & \multicolumn{5}{c|}{Sequential ($q$)} & \multicolumn{7}{c|} {Temporal ($\Delta t$)} & Spatial & \multicolumn{3}{c|}{Spatiotemporal (\# of grids)} \\
		\hline
		Parameters & 1 & 2 & 3 & 4 & 5 & 1h & 2h & 3h & 4h & 6h & 8h & 12h & N/A & $100^2$ & $200^2$ & $300^2$\\
		\hline
		$Acc@1$ & 0.679 & 0.679 & 0.679 & 0.667 & 0.643 & 0.220 & 0.197 & 0.179 & 0.110 & 0.087 & 0.035 & 0.017 & 0.679 & 0.387 & 0.452 & 0.506 \\ 
		\hline
        $Acc@2$ & 0.768 & 0.756 & 0.768 & 0.768 & 0.756 & 0.260 & 0.260 & 0.214 & 0.156 & 0.168 & 0.069 & 0.029 & 0.768 & 0.500 & 0.583 & 0.601 \\ 
        \hline
        $Acc@3$ & 0.815 & 0.839 & 0.821 & 0.815 & 0.810 & 0.289 & 0.295 & 0.277 & 0.208 & 0.220 & 0.092 & 0.035 & 0.815 & 0.554 & 0.643 & 0.649 \\ 
        \hline
        $Acc@4$ & 0.869 & 0.857 & 0.857 & 0.845 & 0.827 & 0.335 & 0.329 & 0.335 & 0.266 & 0.237 & 0.121 & 0.052 & 0.869 & 0.589 & 0.679 & 0.679 \\ 
        \hline
        $Acc@5$ & 0.881 & 0.857 & 0.863 & 0.845 & 0.833 & 0.382 & 0.358 & 0.353 & 0.295 & 0.272 & 0.179 & 0.069 & 0.881 & 0.619 & 0.708 & 0.696 \\ 
        \hline
	\end{tabular}
\end{table*}

\begin{table*}[htb]
    \vspace{-2mm}
	\centering
	\scriptsize
	\renewcommand\arraystretch{1.2}
	\caption{Effectiveness of signature reduction ($|D|=3000$).}
	\label{tab:accuracy_reduction}
	\begin{tabular}{|p{1.4cm}<{\centering}|c|c|c|c|c|c|c|c|c|c|c|c|c|c|c|c|}
		\hline
		Methods & \multicolumn{5}{c|}{PCA} & \multicolumn{5}{c|}{LSH} & \multicolumn{5}{c|}{CUT} & Original \\
		\hline
		$m$ & 10 & 50 & 100 & 500 & 1000 & 10 & 50 & 100 & 500 & 1000 & 10 & 50 & 100 & 500 & 1000 & 160,000 \\
		\hline
		$Acc@1$ & 0.007 & 0.050 & 0.113 & 0.542 & 0.697 & 0.046 & 0.476 & 0.638 & 0.795 & 0.824 & 0.806 & 0.827 & 0.831 & 0.836 & 0.838 & 0.855 \\
		\hline
		$Acc@2$ & 0.012 & 0.088 & 0.187 & 0.686 & 0.801 & 0.079 & 0.542 & 0.705 & 0.847 & 0.870 & 0.866 & 0.877 & 0.880 & 0.885 & 0.886 & 0.904 \\
		\hline
		$Acc@3$ & 0.018 & 0.123 & 0.243 & 0.765 & 0.846 & 0.097 & 0.577 & 0.731 & 0.872 & 0.893 & 0.893 & 0.903 & 0.907 & 0.913 & 0.916 & 0.928 \\
		\hline
		$Acc@4$ & 0.023 & 0.150 & 0.289 & 0.809 & 0.875 & 0.118 & 0.597 & 0.748 & 0.891 & 0.912 & 0.906 & 0.919& 0.920 & 0.928 & 0.929 & 0.940 \\
		\hline
		$Acc@5$ & 0.031 & 0.176 & 0.333 & 0.835 & 0.892 & 0.130 & 0.617 & 0.760 & 0.900 & 0.924 & 0.917 & 0.929 & 0.930 & 0.937 & 0.939 & 0.948 \\
		\hline
	\end{tabular}
\end{table*}

\subsection{Effectiveness}
\label{subsec:effective}
\subsubsection{Signature Representations}
\label{subsubsec:representation}
Table \ref{tab:accuracy_representation} shows the accuracy of moving object linking using sequential, temporal, spatial and spatiotemporal signatures respectively on the dataset $Q$ and $D$ with 3000 randomly sampled taxis. We observe that the accuracy of sequential signatures decreases with subsequence length $q$. Interestingly, the sequential signature is the most effective when $q=1$ which actually corresponds to a geo-spatial point. This reflects that the sequential feature is not helpful for object linking, despite producing strong empirical results for other applications such as location prediction. 
%Note that the accuracy of $1$-gram is inferior to the spatial signature, as we consider commonality only in the generalized Jaccard similarity. 
Temporal signature performs much worse than the other two counterparts, since it simply extracts a coarse distribution of an object's temporal moving behavior. Although the accuracy increases gradually when a finer granularity (i.e., smaller $\Delta t$) is used, the improvement is still insignificant. TF-IDF-based spatial signature is quite effective in modeling an object's traveling behavior, which achieves a 85.5\% accuracy when linking objects to their top-1 nearest neighbors, and the accuracy keeps enhancing when $k$ rises. This verifies the importance of spatial features in moving object modeling and linking. As for the spatiotemporal signature, it defeats the temporal signature thanks to the extra spatial information. However, adding temporal information does not improve performance versus a pure spatial signature under any spatial resolution. This is reasonable since important locations in an object's mobility patterns might be visited at different times, making the patterns extremely sparse. 

Table \ref{tab:accuracy_geolife} reports the accuracy of user linking using the four signature representations on the Geolife dataset. We can easily obtain the similar observations as those on the Beijing taxi dataset. In particular, our algorithm can still achieve a very high performance (i.e., 88.1\% of $Acc@5$) of user linking using the Geolife check-in data, although the accuracy is slightly smaller than that of the taxt data (i.e., 94.8\% of $Acc@5$). This is reasonable, since social media users generally check-in at a lower rate than the GPS points collected by taxis and meanwhile they usually check-in at popular locations (e.g., commonly-visited POIs), making both the commonality and the unicity criteria less effective. Moreover, the sequential signature is not sufficiently useful for user linking in the Geolife data, with its accuracy declines when a larger size of sequential pattern is used (i.e., $q$ in the $q$-gram increases from 1 to 5). 

Overall, the TF-IDF-based spatial signature achieves the best linking accuracy on both taxi and Geolife datasets. 
Hence, we only consider the spatial signature in the remaining experiments.

\subsubsection{Signature Reduction}
\label{subsubsec:reduction}
Table \ref{tab:accuracy_reduction} illustrates the accuracy of various signature reduction strategies (i.e., PCA, LSH, and CUT). The original signature contains around 160,000 points, and we reduce it to $m=$ 10, 50, 100, 500, 1000 points respectively. We have the following observations from Table \ref{tab:accuracy_reduction}: 
1) Linking accuracy degrades with $m$ due to information loss when we reduce the dimensionality of the original signature to $m$. 
2) CUT achieves consistently larger accuracy than LSH which in turn outperforms PCA. Such superiority is extremely evident when the dimensionality is small ($m\leq 100$). PCA is a traditional dimension reduction algorithm and has been empirically proven to be inferior to LSH in many applications. Although LSH is widely-adopted for approximate nearest neighbor search in high-dimensional space (e.g., multimedia search, gene expression identification, near-duplicate document detection, etc.), a simpler method that cuts the long tail of the signature vector has been demonstrated to be more effective in object linking. 
3) Our CUT algorithm can successfully reduce the signature dimensionality from 160,000 to 10 at a slight cost of linking accuracy ($<5\%$). This verifies the possibility of identifying a moving object based on several geo-spatial locations in its traveling history, which is consistent with the phenomenon observed in \cite{de2013unique}. We will use reduced signatures obtained by CUT algorithm with $m=10$ by default for the following experiments.

\subsubsection{Quantitative Signature Metrics}
\label{subsubsec:criteria}
To further understand the proposed spatial signature, we also evaluate the respective contributions of the quantitative signature metrics, i.e., commonality and unicity. The results are reported in Table \ref{tab:accuracy_criteria}. We can observe that the TF-based method performs consistently better than the IDF-based counterpart (avg. 3\% and 10\% accuracy improvement for reduced and original signature respectively), which reflects that commonality of a signature is more important than unicity for modeling an individual's movement behavior. Meanwhile, these two criteria perfectly complement each other, resulting in a much higher accuracy of object linking when combining them via the TF-IDF measure. Such a phenomenon is especially notable for the reduced signature (avg. 70\% accuracy improvement) since only the most important locations in the historical trace are preserved in the reduced signature. This verifies that a signature should be both representative (i.e., high commonality) and distinctive (i.e., high unicity) to effectively identify a moving object.

\begin{table}[htb]
    \vspace{-2mm}
	\centering
	\scriptsize
	\renewcommand\arraystretch{1.2}
	\caption{Contributions of commonality and unicity ($|D|=3000$).}
	\label{tab:accuracy_criteria}
	\begin{tabular}{|c|c|c|c|c|c|c|}
		\hline
		Signature & \multicolumn{3}{c|}{Reduced ($m=10$)} & \multicolumn{3}{c|}{Original} \\
		\hline
		Weighting Strategy & TF & IDF & TF-IDF & TF & IDF & TF-IDF \\
		\hline
		$Acc@1$ & 0.119 & 0.085 & 0.806 & 0.721 & 0.592 & 0.855 \\
		\hline
		$Acc@2$ & 0.161 & 0.126 & 0.866 & 0.765 & 0.666 & 0.904 \\
		\hline
		$Acc@3$ & 0.191 & 0.154 & 0.893 & 0.790 & 0.705 & 0.928 \\
		\hline
		$Acc@4$ & 0.214 & 0.173 & 0.906 & 0.810 & 0.735 & 0.940 \\
		\hline
		$Acc@5$ & 0.231 & 0.187 & 0.917 & 0.823 & 0.750 & 0.948 \\
		\hline
	\end{tabular}
\end{table}

\begin{table}[htb]
	\vspace{-2mm}
	\centering
	\scriptsize
	\renewcommand\arraystretch{1.2}
	\caption{Impact of data split methods ($|D|=3000$).}
	\label{tab:split_method}
	\begin{tabular}{|p{1.2cm}<{\centering}|p{0.6cm}<{\centering}|p{0.85cm}<{\centering}|p{0.85cm}<{\centering}|p{0.6cm}<{\centering}|p{0.85cm}<{\centering}|p{0.85cm}<{\centering}|}
		\hline
		Signature & \multicolumn{3}{c|}{Reduced ($m=10$)} & \multicolumn{3}{c|}{Original} \\
		\hline
		Methods & Serial & Random & Weekday & Serial & Random & Weekday \\
		\hline
		$Acc@1$ & 0.839 & 0.900 & 0.842 & 0.930 & 0.949 & 0.905 \\
		\hline
		$Acc@2$ & 0.903 & 0.943 & 0.904 & 0.958 & 0.972 & 0.947 \\
		\hline
		$Acc@3$ & 0.926 & 0.957 & 0.928 & 0.966 & 0.980 & 0.955\\
		\hline
		$Acc@4$ & 0.936 & 0.964 & 0.940 & 0.972 & 0.987 & 0.965\\
		\hline
		$Acc@5$ & 0.943 & 0.970 & 0.948 & 0.976 & 0.992 & 0.971\\
		\hline
	\end{tabular}
\end{table}

\subsubsection{Data Split Strategies}	
In this work, we divide the original dataset into two parts $Q$ and $D$. Each taxi exists in both datasets, and its trace is divided into two parts accordingly. In particular, we group the trajectories by their dates and assign trajectory groups to $Q$ and $D$ using the following four split methods:
\begin{itemize}
	\item Interleaved: Alternately assign groups to $Q$ and $D$. That is, the trajectories of odd days are assigned to $Q$ and the trajectories of even days are assigned to $D$.
	\item Random: Randomly sample 15 groups and assign to $Q$. The remaining groups are assigned to $D$.
	\item Serial: The first 15 groups are assigned to $Q$ and the remaining 15 groups are assigned to $D$.
	\item Weekday-Weekend: The trajectories of the weekend are assigned to $Q$ and the other trajectories are assigned to $D$.
\end{itemize}
The interleaved datasets are used by default for most of the experiments in this work, to reduce the impact of temporal variation in the object moving behaviors. In addition to these results, Table \ref{tab:split_method} also reports the accuracy of object linking on datasets obtained by other split methods: serial, random, and weekday-weekend. They all outperform the interleaved split in Table \ref{tab:accuracy_size} as they maintain daily continuity in traces. Serially dividing traces obtains a satisfactory linking result, indicating that human mobility patterns are relatively stable during a short time period. Although people might have different traveling preference on weekdays and weekends, our approaches can still discover effective signatures to distinguish different individuals.

\subsubsection{Accuracy and Robustness}
\label{subsubsec:accuracy}
Here we evaluate the robustness of our linking algorithm and the performance of the two optimization schemes (i.e., re-ranking (RR) and stable marriage (SM)). Table \ref{tab:accuracy_size} reports the linking accuracy for different dataset sizes (i.e., number of objects). The accuracy is very high for all datasets, which clearly demonstrates the stability of our signature design.

\begin{table}[ht]
	\vspace{-2mm}
	\centering
	\scriptsize
	\renewcommand\arraystretch{1.2}
	\caption{Effectiveness on different dataset size.}
	\label{tab:accuracy_size}
	\begin{tabular}{|c|p{0.48cm}<{\centering}|p{0.48cm}<{\centering}|p{0.48cm}<{\centering}|p{0.48cm}<{\centering}|p{0.48cm}<{\centering}|p{0.48cm}<{\centering}|p{0.48cm}<{\centering}|p{0.48cm}<{\centering}|}
		\hline
		Signature & \multicolumn{4}{c|}{Reduced ($m=10$)} & \multicolumn{4}{c|}{Original} \\
		\hline
		$|D|$ & 3000 & 6000 & 9000 & 12000 & 3000 & 6000 & 9000 & 12000 \\
		\hline
		$Acc@1$ & 0.806 & 0.767 & 0.755 & 0.754 & 0.855 & 0.831 & 0.825 & 0.829 \\
		\hline
		$Acc@2$ & 0.866 & 0.837 & 0.825 & 0.825 & 0.904 & 0.879 & 0.874 & 0.874 \\
		\hline
		$Acc@3$ & 0.893 & 0.867 & 0.860 & 0.858 & 0.928 & 0.900 & 0.894 & 0.896 \\
		\hline
		$Acc@4$ & 0.906 & 0.884 & 0.877 & 0.877 & 0.940 & 0.914 & 0.908 & 0.910 \\
		\hline
		$Acc@5$ & 0.917 & 0.893 & 0.890 & 0.888 & 0.948 & 0.922 & 0.916 & 0.917 \\
		\hline
	\end{tabular}
\end{table}

\begin{table}[htb]
    \vspace{-2mm}
	\centering
	\scriptsize
	\renewcommand\arraystretch{1.2}
	\caption{Effectiveness of two optimization schemes on different dataset sizes.}
	\label{tab:optimisation_acc}
	\begin{tabular}{|p{1.8cm}<{\centering}|p{0.6cm}<{\centering}|p{0.6cm}<{\centering}|p{0.6cm}<{\centering}|p{0.6cm}<{\centering}|p{0.9cm}<{\centering}|c|}
		\hline
		Scheme & {\texttimes} & \multicolumn{4}{c|}{RR} & {SM}\\
		\hline
		$m$ & 10 & 50 & 100 & 500 & original & 10 \\
		\hline
		$|D|=3000$ & 0.806 & 0.823 & 0.828 & 0.834 & 0.841 & 0.845 \\
		\hline
		$|D|=6000$ & 0.767 & 0.792 & 0.797 & 0.805 & 0.814 & 0.813 \\
		\hline
		$|D|=9000$ & 0.755 & 0.789 & 0.794 & 0.799 & 0.811 & 0.802 \\
		\hline
		$|D|=12000$ & 0.754 & 0.791 & 0.798 & 0.803 & 0.815 & 0.803 \\
		\hline
	\end{tabular}
\end{table}

\begin{table}[htb]
    \vspace{-2mm}
	\centering
	\scriptsize
	\renewcommand\arraystretch{1.2}
	\caption{Extra time cost (s) of optimization schemes on different dataset sizes.}
	\label{tab:optimisation_time}
	\begin{tabular}{|p{1.5cm}<{\centering}|p{0.6cm}<{\centering}|p{0.6cm}<{\centering}|p{0.6cm}<{\centering}|p{0.6cm}<{\centering}|p{0.9cm}<{\centering}|c|}
		\hline
		Scheme & {\texttimes} & \multicolumn{4}{c|}{RR} & {SM}\\
		\hline
		$m$ & 10 & 50 & 100 & 500 & original & 10 \\
		\hline
		$|D|=3000$ & 0.164 & 0.026 & 0.049 & 0.255 & 13.886 & 0.190 \\
		\hline
		$|D|=6000$ & 0.464 & 0.057 & 0.108 & 0.499 & 28.684 & 0.497  \\
		\hline
		$|D|=9000$ & 0.898 & 0.118 & 0.204 & 1.131 & 40.732 & 0.972 \\
		\hline
		$|D|=12000$ & 1.738 & 0.141 & 0.246 & 1.576 & 52.258 & 1.804 \\
		\hline
	\end{tabular}
\end{table}

Table \ref{tab:optimisation_acc} reports the accuracy improvement using RR and SM for $Acc@1$. For RR, we first use $10$-dim signatures to quickly find the top-$k$ candidates for each object, then the $k$ candidates are re-ranked using larger signatures of sizes $50, 100, 500$ and the full-size for a more accurate similarity comparison.  Algorithm \ref{algo:marriage} is used to apply stable marriage reshuffle ($k=5$ for all experiments here). From Table \ref{tab:optimisation_acc} we observe that two strategies are very effective. Compared with the result of only using $10$-dim signatures, the SM process leads to a $4.5\%$ increase of linking accuracy on average, while the RR improvement depends on different sizes of signatures that are used (ranging from $1.7\%$ to $6.1\%$). The additional time costs for applying RR and SM are shown in Table \ref{tab:optimisation_time}. To put the cost for re-ranking into context, full-sized signatures can have an average of $160,000$ points and the re-ranking stage with full-sized signatures takes much longer time with little accuracy gain (as shown above). The revised SM process is slightly superior over the RR method, since it achieves a larger accuracy improvement with a comparable time cost ($m=500$ for RR).
%For the revised SM process, it consumes more time than RR under most settings while the extra costs are still very acceptable. 
Overall, it is worthwhile to spend less than two more seconds for a nearly $5\%$ accuracy increase (in the case of $|D|=12000$).

Furthermore, we examine whether the combination of RR and SM can achieve even better accuracy. Table \ref{tab:optimisation_combine} shows that applying SM on the sets of RR-improved top-$5$ candidates maintains a significant improvement of about $4\%$ for $Acc@1$. As $k$ increases, the accuracy gain diminishes. Therefore, combining RR and SM is a good strategy to improve linking accuracy, especially when the top-$1$ candidate is of the main interest.

\begin{table}[htb]
    \vspace{-2mm}
	\centering
	\scriptsize
	\renewcommand\arraystretch{1.2}
	\caption{Accuracy using RR and SM with different $m$ ($|D|=3000$).}
	\label{tab:optimisation_combine}
	\begin{tabular}{|p{0.8cm}<{\centering}|p{1.1cm}<{\centering}|p{0.5cm}<{\centering}|p{0.5cm}<{\centering}|p{0.5cm}<{\centering}|p{0.5cm}<{\centering}|p{0.5cm}<{\centering}|p{0.5cm}<{\centering}|p{0.5cm}<{\centering}|p{0.5cm}<{\centering}|}
		\hline
		Scheme & {SM} & \multicolumn{3}{c|}{RR} & \multicolumn{3}{c|}{Combination} \\ \hline
		$m$ & 10 & 50 & 100 & 500 & 50 & 100 & 500 \\ \hline
        $Acc@1$ & 0.845 & 0.823 & 0.828 & 0.834 &  0.864 & 0.869 & 0.874  \\ \hline
        $Acc@2$ & 0.883 & 0.873 & 0.875 & 0.876 &  0.898 & 0.901 & 0.902  \\ \hline
        $Acc@3$ & 0.904 & 0.898 & 0.896 & 0.901 &  0.908 & 0.907 & 0.917  \\ \hline
        $Acc@4$ & 0.915 & 0.909 & 0.910 & 0.911 &  0.916 & 0.919 & 0.921  \\ \hline
        $Acc@5$ & 0.923 & 0.917 & 0.917 & 0.917 &  0.923 & 0.925 & 0.926  \\ \hline
	\end{tabular}
\end{table}

\subsubsection{Linkability and Signature Sensitivity}
\label{subsubsec:linkability}
\begin{figure}[htb]
\centering
\begin{tabular}{rl}
\subfigure[$d$ query days vs $15$ days] {
\resizebox{0.22\textwidth}{!}{%
\begin{tikzpicture}

    \begin{axis}[
    %width=0.5\textwidth,
    height=5cm,
    ymajorgrids,
    ymin=0.3,
    %ytick scale label code/.code={$\times$bn\EUR{}},
    legend pos= south east
    ]
        \addplot table[x=days,y=Acc@1] {data_query.csv};\addlegendentry{\tiny Acc@1}
        \addplot table[x=days,y=Acc@2] {data_query.csv};\addlegendentry{\tiny Acc@2}
        \addplot table[x=days,y=Acc@3] {data_query.csv};\addlegendentry{\tiny Acc@3}
        \addplot table[x=days,y=Acc@4] {data_query.csv};\addlegendentry{\tiny Acc@4}
        \addplot table[x=days,y=Acc@5] {data_query.csv};\addlegendentry{\tiny Acc@5}

    \end{axis}
\end{tikzpicture}
}
} &

\subfigure[$d$ days vs $d$ days]{
\resizebox{0.22\textwidth}{!}{%
\begin{tikzpicture}
    \begin{axis}[
    %width=0.5\textwidth,
    height=5cm,
    ymajorgrids,
    ymin=0.3,
    %ytick scale label code/.code={$\times$bn\EUR{}},
    legend pos= south east
    ]
        \addplot table[x=days,y=Acc@1] {data.csv};\addlegendentry{\tiny Acc@1}
        \addplot table[x=days,y=Acc@2] {data.csv};\addlegendentry{\tiny Acc@2}
        \addplot table[x=days,y=Acc@3] {data.csv};\addlegendentry{\tiny Acc@3}
        \addplot table[x=days,y=Acc@4] {data.csv};\addlegendentry{\tiny Acc@4}
        \addplot table[x=days,y=Acc@5] {data.csv};\addlegendentry{\tiny Acc@5}

    \end{axis}
\end{tikzpicture}
}
}
\end{tabular}

\caption{Effectiveness on different information size ($|D|=3000,m=10$).}
\label{fig:accuracy_days}
\end{figure}
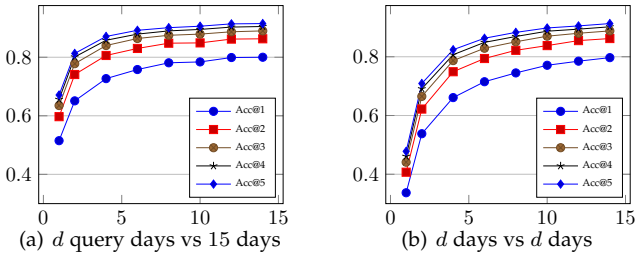

Now, we examine the impact of information size on linkability: how much data is needed for reliable linking? The trajectories in query set $Q$ are limited to $d$ days of data, from 1 to 15 days (full data); Linking is performed with a constant full 15-day dataset (Figure \ref{fig:accuracy_days}-a) and a similarly limited data set using $d$ days of trajectories (Figure \ref{fig:accuracy_days}-b). The $d$ days are chosen at random and the average accuracy is reported for multiple runs. As depicted, the linking accuracy increases significantly with the amount of data, and the accuracy is significantly higher when searching on the full dataset. The accuracy becomes more robust when $d\geq 7$ in both cases. In other words, the signatures become sufficient to guarantee high linking accuracy with only one week's data.
%Hence, the effective and robust signatures are supposed to be extracted based on at least one-week data, so that the linking performance can be guaranteed confidently.

While the signatures proposed in this paper can achieve highly accurate linking, it is necessary to understand the importance of the points in the signature closure (Section \ref{subsec:privacy}) by examining if objects can still be linked after those points in the signatures are removed. Table~\ref{tab:closure} reports the linking accuracy using the $10$-dim signatures on the Beijing taxi dataset.
%Our experimental results strongly demonstrate the power of TF-IDF weighted spatial signatures for linkability protection.
In the 1st round, the signatures are generated using the entire dataset; in the 2nd round, a new signature for each object is generated using the dataset with the points in the previous signatures removed; and so on. In each iteration, we test both the basic ``$Acc@1$" and the ``Best $Acc@5$" which is derived from the combination of re-ranking and stable marriage with $500$-dim signatures (i.e., the best linking accuracy we can achieve). In addition, we also measure the data \textit{utility} affected by removing the top-10 points from the signatures by calculating the average percentage of data remaining, as well as the average overlapping of objects' MBRs and the average percentage of cells in a uniform grid covered by the trajectories before and after signature points are removed 
%by calculating the percentage of data remaining, the average percentage of the new MBR for all the points of an object compared with that of the original data, as well as the average percentage of the number of grids they occupy compared with the number of grids they occupy originally 
(we use `small' and `large' grids which are of   $85\times 85$ and $423 \times 423$ $m^2$, respectively). 
From Table \ref{tab:closure} we observe a dramatic decrease of linking accuracy each time when the points in the signatures from the previous round are removed, while the data utility indicators remain to be very high. After removing the signatures for just one round (i.e., removing 10 points for each object which has an average of 30,000 points originally), the best linking accuracy drops from an initial $92.6\%$ to an unusable $45\%$. 

A similar experiment is also conducted on the Geolife dataset, since the privacy issues are usually more important for human mobility patterns rather than taxis. The results of linking accuracy and utility metrics are reported in Table \ref{tab:closure_geolife}. It can be observed that our proposed ``signature closure'' algorithm still works well for the Geolife dataset. After iteratively removing the top-$10$ signature points for five times, the linking accuracy under our best configuration (i.e., the combination of re-ranking and stable marriage) decreases to around 10\%, making the users less linkable based on the modified trajectories. Moreover, the unicity of the dataset only slightly decreased after several rounds of signature suppression. 

Results in Table \ref{tab:closure} and Table \ref{tab:closure_geolife} clearly illustrate that the signature points of an object generated using our approach are highly representative and discriminative to the object. These points are suitable as signature representations for object linking. Conversely, it is a very promising approach to protect trajectory data by removing a small number of points (i.e., the points in the signature closure, or the first a few rounds of it) such that the re-identification attacks can be prevented to a large extent. Trajectory privacy protection is beyond the scope of this paper, but we present some interesting and exciting insights here for new ways to protect privacy of trajectory data. By the suppression or anonymization of a small number of points in the signature closure, we can achieve strong results for preventing link-based attacks at a very limited cost of data utility. We will further investigate signature closure in the future.

\begin{table}[htb]
	\centering
	\scriptsize
	\renewcommand\arraystretch{1.2}
	\caption{Signature sensitivity ($|D|=3000, m=10$).}
	\label{tab:closure}
	\begin{tabular}{|p{0.6cm}<{\centering}|c|c|c|c|c|c|}
    \hline
    \multirow{2}{*}{Round} & \multicolumn{2}{c|}{Linking Accuracy} & \multirow{2}{*}{\makecell[c]{Data\\ Remain}} & \multirow{2}{*}{\makecell[c]{MBR \\ Overlap}} & \multicolumn{2}{c|}{Grid Coverage} \\ \cline{2-3}\cline{6-7}
    & $Acc@1$ & Best $Acc@5$ & & & Large % $423^2$ m 
    & Small % $85^2 m^2$ 
    \\ \hline
    1  & 0.806 & 0.926 & 0.978 & 0.999 & 0.994 & 0.961 \\ \hline
    2  & 0.140 & 0.450 & 0.969 & 0.999 & 0.989 & 0.944 \\ \hline
    3  & 0.050 & 0.187 & 0.960 & 0.999 & 0.984 & 0.928 \\ \hline
    4  & 0.028 & 0.114 & 0.953 & 0.998 & 0.980 & 0.913 \\ \hline
    5  & 0.020 & 0.078 & 0.947 & 0.998 & 0.975 & 0.898 \\ \hline
%    6  & 0.013 & 0.079 & 0.941 & 0.997 & 0.971 & 0.884 \\ \hline
%    7  & 0.006  & 0.935 & 0.997 & 0.967 & 0.871 \\ \hline
%    8  & 0.010  & 0.930 & 0.997 & 0.963 & 0.857 \\ \hline
\end{tabular}
\end{table}

\begin{table}[htb]
	\centering
	\scriptsize
	\renewcommand\arraystretch{1.2}
	\caption{Signature sensitivity on Geolife ($|D|=175, m=10$).}
	\label{tab:closure_geolife}
	\begin{tabular}{|p{0.6cm}<{\centering}|c|c|c|c|c|c|}
    \hline
    \multirow{2}{*}{Round} & \multicolumn{2}{c|}{Linking Accuracy} & \multirow{2}{*}{\makecell[c]{Data\\ Remain}} & \multirow{2}{*}{\makecell[c]{MBR \\ Overlap}} & \multicolumn{2}{c|}{Grid Coverage} \\ \cline{2-3}\cline{6-7}
    & $Acc@1$ & Best $Acc@5$ & & & Large % $423^2$ m 
    & Small % $85^2 m^2$ 
    \\ \hline
    1 & 0.655 & 0.839 & 0.928 & 0.781 & 0.921 & 0.879 \\ \hline
    2 & 0.286 & 0.524 & 0.888 & 0.700 & 0.879 & 0.830 \\ \hline
    3 & 0.107 & 0.286 & 0.856 & 0.670 & 0.844 & 0.786 \\ \hline
    4 & 0.077 & 0.179 & 0.826 & 0.645 & 0.811 & 0.751 \\ \hline
    5 & 0.048 & 0.125 & 0.797 & 0.621 & 0.781 & 0.719 \\ \hline
\end{tabular}
\end{table}

\subsection{Efficiency}
\label{subsec:efficiency}
\subsubsection{Object Linking Algorithms}
\label{subsubsec:baselines}
Table \ref{tab:efficiency} reports the efficiency of object linking algorithms on datasets of various sizes. L2AP takes the longest time for object linking (outperformed even by the naive linear scan), though it is considered as an efficient approach to cosine similarity search. It is designed for a really high-dimensional space (e.g., million-level), hence is unsuitable for handling reduced signatures as it wastes a large amount of time for excessive bound checking and index construction. LSH is also proposed for speeding up $k$-NN search in a high dimension (e.g., multimedia data). It only achieves a minor performance improvement compared to the linear scan when signature dimension is reduced to $m=10$. Overall, our approach works very well by transforming the linking problem to 2D space and utilize R-tree and spatial overlapping for search pruning, as illustrated in 2D R-tree and WR-tree. More importantly, WR-tree outperforms 2D R-tree by a large margin. This verifies the effectiveness of both pruning strategies introduced in this work, namely spatial overlapping and signature similarity. Table \ref{tab:efficiency} also shows the impact of dataset size $|D|$. As expected, the linking time rises with $|D|$, but when the WR-tree is used the increase is quite small ($\approx 0.6s$ when $|D|$ ranges from 3000 to 12000).

\begin{table}[htb]
    \vspace{-2mm}
	\centering
	\scriptsize
	\renewcommand\arraystretch{1.2}
	\caption{ Total time cost (s) of different linking algorithms ($m=10$, $k=1$).}
	\label{tab:efficiency}
	\begin{tabular}{|p{1.2cm}<{\centering}|p{0.9cm}<{\centering}|p{0.9cm}<{\centering}|p{0.9cm}<{\centering}|p{0.9cm}<{\centering}|p{1.2cm}<{\centering}|}
		\hline
		$|D|$ & Linear & L2AP & LSH & R-tree & WR-tree \\
		\hline
		$3000$ & 2.269 & 3.090 & 1.769 & 0.364 & 0.059 \\
		\hline
		$6000$ & 8.182 & 14.557 & 6.652 & 1.518 & 0.220 \\
		\hline
		$9000$ & 19.733 & 36.541 & 15.642 & 3.597 & 0.387 \\
		\hline
		$12000$ & 27.183 & 70.440 & 38.131 & 7.206 & 0.678 \\
		\hline
	\end{tabular}
\end{table}

\begin{comment}
\begin{figure}[htp]
    \vspace{-2mm}
    \centering
    \caption{Total time cost (s) of different linking algorithms ($m=10$, $k=1$).}
    \label{fig:efficiency}
    \begin{tikzpicture}
        
    \begin{axis}[
        %axis lines = left,
        xlabel = {$|D|$},
        %ylabel = {$Time~cost~(s)$},
        xtick=data,
        width=9cm, height=9cm, 
        %tick align=outside,
        legend pos=north west]
        
        \addplot[mark=square,draw=green] coordinates {(3000,2.269) (6000,8.182) (9000,19.733) (12000,27.183)};
        \addlegendentry{Linear}
        \addplot[mark=o,draw=gray] coordinates {(3000,3.090) (6000,14.557) (9000,36.541) (12000,70.440)};
        \addlegendentry{L2AP}
        \addplot[mark=x,draw=brown] coordinates {(3000,1.769) (6000,6.652) (9000,15.642) (12000,38.131)};
        \addlegendentry{LSH}
        \addplot[mark=+, draw=blue] coordinates {(3000,0.364) (6000,1.518) (9000,3.597) (12000,7.206)};
        \addlegendentry{2D R-tree}
        \addplot[mark=star,draw=red] coordinates {(3000,0.068) (6000,0.246) (9000,0.338) (12000,0.651)};
        \addlegendentry{WR-tree}
    \end{axis}
    \end{tikzpicture}
\end{figure}
\end{comment}

\subsubsection{WR-Tree Construction and Update}
\label{subsubsec:index}
Recall that we design a new criterion to optimize WR-tree (i.e., signature enlargement), which motivates the introduction of bulk-loading and incremental algorithms for WR-tree construction and update, respectively. Table \ref{tab:construction} examines whether the efficiency of WR-tree can be further improved by considering signature enlargement in index construction. We can see that both the linking cost and index size are slightly reduced when the aggregated signature sizes at internal tree nodes are minimized by maximizing the number of common points between signatures. Such improvement increases with the growth of the signature size $m$. This is because larger signatures will increase not only the cost of similarity calculation but also the possibility of spatial overlapping (recall that our CUT dimension reduction has a natural side effect of ``spatial shrinking''). Table \ref{tab:construction} also empirically demonstrates the superiority of the bulk-loading index construction approach compared to the insertion-based algorithm adopted in \cite{jin2019moving}, with the linking efficiency improved at least two-fold thanks to a global optimization of the WR-tree structure. Although the index construction time also increases if signature enlargement is incorporated into the bulk-loading algorithm, it is still acceptable since the index is usually constructed only once.

We then examine the efficiency of WR-tree updates. The base WR-tree is built for the dataset with 9,000 objects. The index update costs and the linking efficiency are studied by repeatedly inserting 500 new objects one by one. %Again, here we consider only appending newly arrived objects' signatures into the existing tree. The temporal evolution or other types of update of signatures that belong to those objects that have been indexed will be a promising future work. 
% As discussed in Section \ref{subsec:update}, the new criterion of signature enlargement  considers selecting a proper sub-tree for adding the new object to the index.
%to minimize the signature size of internal nodes (by maximizing the number of common points between signatures).
Table \ref{tab:update} presents the cost to rebuild the WR-tree using the bulk-loading method, the average cost of each update (i.e., appending one object), and the total time of object linking. As expected, the update cost is extremely small compared with the cost of rebuilding the WR-tree from scratch. It verifies the effectiveness of our proposed incremental tree maintenance. A slight decrease in both the tree update time and the linking cost can be observed when the new criterion of signature enlargement is considered in index update. Compared with a newly-built tree, the linking cost of the updated tree increases due to the imperfect tree structure caused by the incremental updates. However, WR-tree is still far superior over other linking methods as shown in Table \ref{tab:efficiency} in terms of the overall linking performance. 
%The average cost of each update is much smaller than the index rebuild time. 
Clearly, incremental WR-tree maintenance is effective while a periodic index reconstruction is still necessary. This fact provides a solid base for balancing system throughput and query response time when applying the WR-tree in practice.

\begin{comment}
\begin{table}[htb]
    \vspace{-2mm}
	\centering
	\scriptsize
	\renewcommand\arraystretch{1.2}
	\caption{Total time cost (s) of WR-tree on variable parameters ($|D|=3000$).}
	\label{tab:efficiency_m_k}
	\begin{tabular}{|p{1.1cm}<{\centering}|p{0.9cm}<{\centering}|p{0.9cm}<{\centering}|p{0.9cm}<{\centering}|p{0.9cm}<{\centering}|p{0.9cm}<{\centering}|}
	\hline
	 & $k=1$ & $k=2$ & $k=3$ & $k=4$ & $k=5$ \\ \hline
    $m=10$ & 0.059 & 0.088 & 0.137 & 0.166 & 0.184 \\ \hline
    $m=50$ & 0.508 & 0.691 & 0.808 & 0.892 & 0.997 \\ \hline
    $m=100$ & 1.029 & 1.360 & 1.709 & 1.984 & 2.245 \\ \hline
	\end{tabular}
\end{table}
\end{comment}

\begin{table}[htb]
    \vspace{-2mm}
	\centering
	\scriptsize
	\renewcommand\arraystretch{1.2}
	\caption{Performance of WR-tree construction w.r.t build time (s), linking time (s) and index size (M) ($|D|=3000, k=1$).}
	\label{tab:construction}
	\begin{tabular}{|p{1.8cm}<{\centering}|c|c|c|c|c|c|}
		\hline
		Build Criteria & \multicolumn{3}{c|}{Spatial Factors} & \multicolumn{3}{c|}{\# Common Points} \\
		\hline
		Cost & Build & Link & Size & Build & Link & Size \\
		\hline
		$m=10$  & 0.078 & 0.077 & 3.650   & 1.278  & 0.059 & 3.571  \\ \hline
        $m=50$  & 0.332 & 0.681 & 15.434 & 7.501  & 0.447 & 14.892 \\ \hline
        $m=100$ & 0.604 & 1.759 & 29.067 & 15.694 & 1.232 & 27.642 \\ \hline

	\end{tabular}
\end{table}

\begin{table}[htb]
	\centering
	\scriptsize
	\renewcommand\arraystretch{1.2}
	\caption{Efficiency (s) of the newly built WR-tree and the updated WR-tree with different update criteria ($m=10, k=1$).}
	\label{tab:update}
	\begin{tabular}{|p{0.6cm}<{\centering}|c|c|c|c|c|c|}
		\hline
		 & \multicolumn{2}{c|}{Newly Built} & \multicolumn{2}{c|}{Spatial Factors} & \multicolumn{2}{c|}{\# Common Points} \\
		 \hline
        $|D|$ & Build & Link & Per Update & Link & Per Update & Link \\ \hline
        $9500$  & 9.642  & 0.399 & 0.0099 & 0.425 & 0.0096 & 0.404 \\ \hline
        $10000$ & 9.879  & 0.436 & 0.0107 & 0.477 & 0.0104 & 0.451 \\ \hline
        $10500$ & 10.062 & 0.471 & 0.0115 & 0.512 & 0.0113 & 0.498 \\ \hline
        $11000$ & 10.107 & 0.528 & 0.0135 & 0.623 & 0.0119 & 0.592 \\ \hline
        $11500$ & 10.278 & 0.566 & 0.0157 & 0.695 & 0.0126 & 0.664 \\ \hline
        $12000$ & 10.455 & 0.609 & 0.0175 & 0.824 & 0.0131 & 0.794 \\ \hline
	\end{tabular}
\end{table}

\begin{comment}
\begin{table}[htb]
	\centering
	\scriptsize
	\renewcommand\arraystretch{1.2}
	\caption{Efficiency and Time cost (s) of 1) the newly built WR-tree and 2) the updated WR-tree with different insertion methods ($m=10, k=1$).}
	\label{tab:construction}
	\begin{tabular}{|p{0.8cm}<{\centering}|c|c|c|c|c|c|}
		\hline
		 & \multicolumn{2}{c|}{Newly Built} & \multicolumn{2}{c|}{Spatial Factors} & \multicolumn{2}{c|}{\# Common Points} \\
		\hline
		$|D|$ & Build & Link & Per Update & Link & Per Update & Link \\ 
		\hline
		$9500$ & 0.278 & 0.457 & 0.0156 & 0.631 & 0.0335 & 0.684 \\
		\hline
		$10000$ & 0.317 & 0.504 & 0.0134 & 0.729 & 0.0332 & 0.826 \\
		\hline
		$10500$ & 0.325 & 0.544 & 0.0131 & 0.787 & 0.0332 & 0.943 \\
		\hline
		$11000$ & 0.336 & 0.572 & 0.0130 & 0.850 & 0.0336 & 1.061 \\
		\hline
		$11500$ & 0.341 & 0.586 & 0.0136 & 0.944 & 0.0338 & 1.144 \\
		\hline
		$12000$ & 0.349 & 0.614 & 0.0148 & 1.111 & 0.0320 & 1.251 \\
		\hline
	\end{tabular}
\end{table}
\end{comment}

\section{Related Work}
\label{sec:relatedwork}
In this section, we briefly summarize the existing work in several research areas that are related to our work, including trajectory pattern mining, cosine similarity search, spatial index, and trajectory privacy protection.

\begin{comment}
\subsection{Entity linking}
Entity linking is the task of determining the identity of an entity mentioned in a data source by either linking it to an entry in a knowledge base [?] or linking together multiple mentions of the same entity in different data sources [?]. It is a fundamental step to make data useful, as information about an entity can be enriched from different sources about the same entity if they can be reliably and efficiently linked together. This problem has been extensively studied for text data (where a named entity is represented as a text string). For example, once “Michael Jordan” is recognized as a named entity, it can be linked, using different techniques such as the contextual information of its mentions in a document, to a knowledge base to disambiguate its meaning as, for example, a basketball player, a machine learning expert or a brand of sports shoes. There exist many reliable and efficient solutions that have been used as the foundation of data integration to improve the performance of information retrieval [?], to support data augmentation [?] and to enable data quality management in general [?]. Similar demands can be found for other types of data such as video data (an object such as a person or vehicle can be linked together cross different surveillance cameras based on their extracted visual features [?]).
\end{comment}

{\em Trajectory pattern mining:}
The existing works on trajectory pattern mining can be divided into three categories based on different definitions of patterns. \textit{Sequential} pattern mining \cite{Cao2005}\cite{Giannotti2007} identifies a common sequence of locations traveled by a certain number of objects within a similar timeslot. Another research branch \cite{Cao2007}\cite{Li2010}\cite{Li2012} tries to discover \textit{periodic} activity patterns from the movement history, which are then used for predicting the future behavior of a moving object. Other works attempt to detect \textit{collective} moving patterns, namely, a group of objects that always travel together for a certain period, such as convoy \cite{Jeung2008Convoy}, swarm \cite{Li2010Swarm}, traveling companion \cite{Tang2012Companion}, and gathering \cite{Zheng2013Gather}, etc.  Unlike the described traditional research, our work focuses on extracting patterns that are both common and unique to a moving object to support object identification.

{\em Cosine similarity search:}
Given two sets of weighted vectors and a threshold, the cosine similarity search finds all vector pairs whose cosine similarity exceeds the threshold. Existing work mainly focuses on quickly locating all necessary candidate pairs by checking only a few elements at the beginning of the vectors, which is well-known as the \textit{pruning-verification} framework. The AllPairs \cite{Bayardo2007} algorithm avoids computing similarity for unpromising vectors by exploiting a dynamically-constructed inverted index. Extensions to the AllPairs algorithm, such as APT \cite{Awekar2009}, MMJoin \cite{Lee2010} and L2AP \cite{Anastasiu2014}, further improve the efficiency by introducing tighter similarity bounds. Alternatives also exist that find approximate answers for cosine similarity search using locality sensitive hashing (LSH) \cite{Indyk1998}\cite{Gionis1999}. However, the semantics of vector elements (i.e., the spatial features of GPS points) are ignored in these methods, which limits their efficiency in our problem setting. We show in this work how the geographical information in object signatures can be exploited to improve object linking efficiency.

{\em Spatial indexing:}
Spatial index is designed for organizing spatial objects and optimizing a wide range of spatial queries. Various indexing structures have been proposed in the literature, including Quadtree \cite{Finkel1974}, R-tree \cite{Guttman1984R}\cite{Sellis1987RPlus}\cite{Beckmann1990RStar}\cite{Kamel1994HilbertR}\cite{Berchtold1996X}\cite{Arge2008PR}, k-d tree \cite{Bentley1975KD1}\cite{Bentley1979KD2}\cite{Robinson1981KDB}, hB-tree \cite{Lomet1990HB}\cite{Evangelidis1997HBPi}, etc. Specifically, R-tree \cite{Guttman1984R} is a dynamic and balanced tree structure that organizes minimum bounding rectangles (MBRs) of spatial objects into leaf nodes and groups nearby MBRs into internal nodes. The search algorithm determines whether or not to search inside a subtree by checking MBRs. A majority of nodes are pruned in this way, reducing I/O cost. The construction method of R-tree, in particular how to group MBRs into intermediate nodes, is indispensable to its search performance in practice. Most of the R-tree variants, e.g., R+ tree \cite{Sellis1987RPlus}, R*-tree \cite{Beckmann1990RStar}, Hilbert R-tree \cite{Kamel1994HilbertR}, X-tree \cite{Berchtold1996X}, PR-tree \cite{Arge2008PR}, etc., target at improving the merging strategy. In this work, we extend the R-tree structure to combine both spatial and weight information of each point for more efficient moving object linking.

{\em Trajectory privacy protection:}
Privacy-preserving trajectory publication aims to release individuals' moving trajectories without leaking their sensitive information. It has been extensively studied in the literature, adapting various  classic data privacy protection models to trajectory data. $k$-anonymity algorithms (e.g., W4M \cite{abul2010anonymization} and GLOVE \cite{gramaglia2015hiding}) cluster and merge trajectories such that a trajectory is indistinguishable with at least other $k$-1 trajectories. $l$-diversity and $t$-closeness \cite{tu2018protecting} further extend $k$-anonymity to prevent semantic attack by considering sensitive attribute value distributions in trajectories. Differential privacy (e.g., DPT \cite{he2015dpt}) and plausible deniability \cite{bindschaedler2016synthesizing} insert random noises into trajectories, ensuring that the presence of an entity or trajectory in a dataset can only be observed for a controlled amount of certainty. Besides, some ad-hoc privacy models, such as dummy \cite{liu2019dummy} and mix-zone \cite{liu2012traffic}, have also been applied to anonymize sensitive information for trajectory data. In this work, we show that the TF-IDF-weighted signatures are highly effective for protecting trajectory privacy, especially the linkability attack, through the signature closure concept. This should potentially be a promising solution for trajectory anonymization, and could be further investigated in the future work.

\section{Conclusion}
\label{sec:conclusion}
In this paper, we have studied the problem of spatiotemporal entity linking based on trajectories. Different signature representation strategies have been examined for their ability to capture the unique characteristics of trajectory data. The linking problem is formalized as a $k$-NN query based on signature similarity. A comprehensive suite of techniques have been developed, including signature reduction and WR-tree indexing with update support. Two optimization schemes (i.e., re-ranking and stable marriage) are introduced to enhance the linking accuracy. An empirical study on a large taxi dataset demonstrates significantly better accuracy and efficiency of our approach than the state-of-the-art methods. In the future, we will extend our research to heterogeneous datasets over a long period of time to support cross-domain spatiotemporal entity linking. The location privacy projection issue will also be further investigated.
\section*{Acknowledgment}
This work was partially supported by the Australian Research Council (Grant No. DP200103650 and LP180100018).

\ifCLASSOPTIONcaptionsoff
  \newpage
\fi

\bibliographystyle{IEEEtran}
\bibliography{oblinking}

\vspace*{-1.2cm}

\begin{IEEEbiography}[{\includegraphics[width=1in,height=1.25in,clip,keepaspectratio]{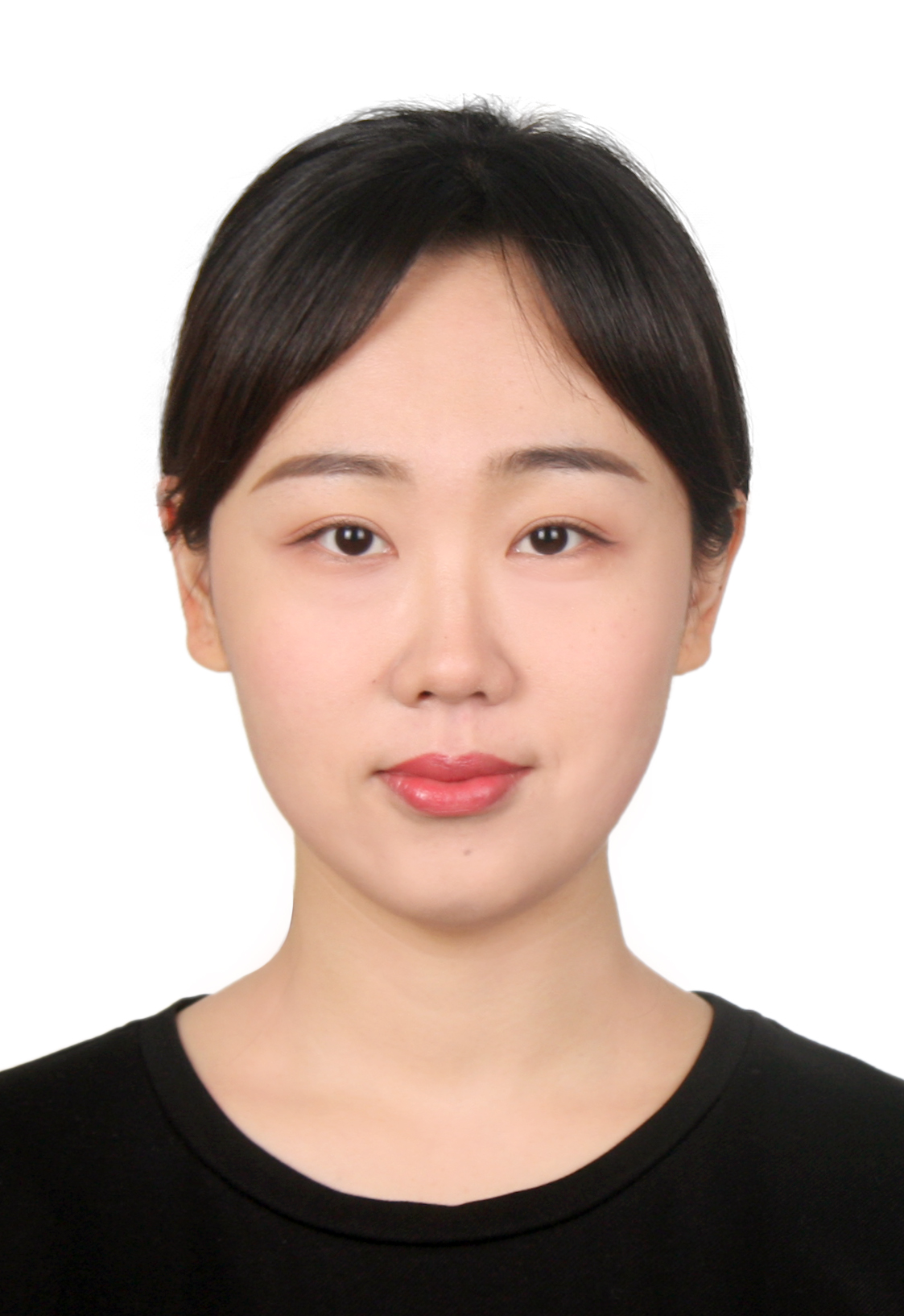}}]{Fengmei Jin}
received her Bachelor of Engineering from Sun Yat-Sen University in 2016 and Master of Engineering from Renmin University of China in 2019. Currently, she is a PhD candidate at The University of Queensland. Her research interests include spatiotemporal databases, pattern mining, and data integration.
\end{IEEEbiography}

\vspace*{-1.3cm}

\begin{IEEEbiography}
[{\includegraphics[width=1in,height=1.25in,clip,keepaspectratio]{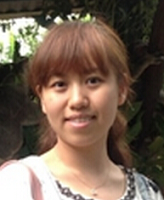}}]{Wen Hua}
is a Lecturer at The University of Queensland. She received her PhD and Bachelor degrees in computer science from Renmin University of China in 2015 and 2010, respectively. Her main research interests include database systems, information extraction, data integration, and spatiotemporal data management. % She has published actively in reputed journals and international conferences and won several Best Paper Awards.
\end{IEEEbiography}

\vspace*{-1.3cm}

\begin{IEEEbiography}[{\includegraphics[width=1in,height=1.25in,clip,keepaspectratio]{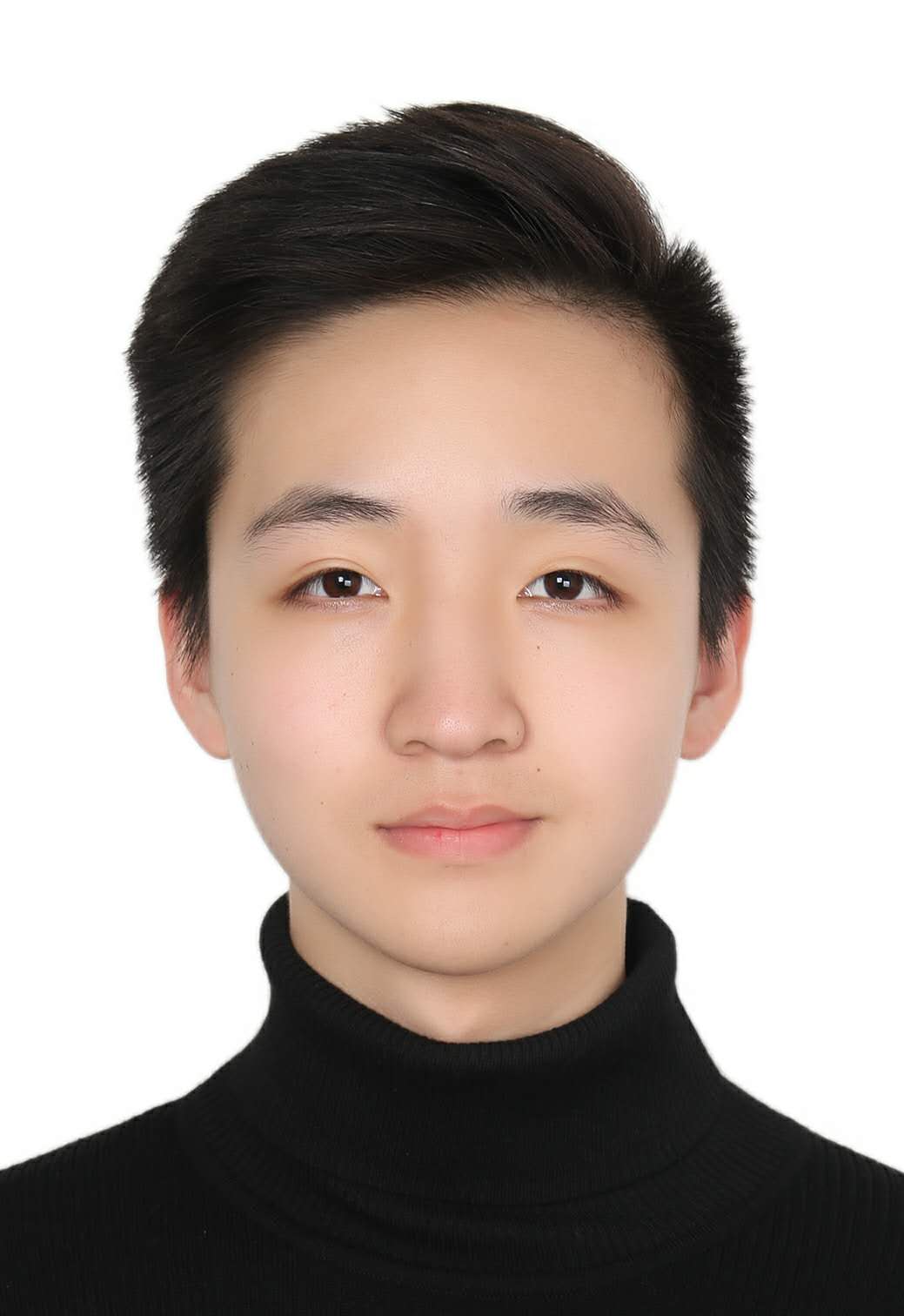}}]{Thomas Zhou}
received his Bachelor of Science (Computer Science) degree with First Class Honors from The University of Queensland in 2019. He is a Research Assistant at The University of Queensland.
%Currently he is a PhD candidate in the Department of Biomedical Engineering, The Chinese University of Hong Kong. He is a recipient of Hong Kong PhD Fellowship Scheme (HKPFS) scholarship.
\end{IEEEbiography}

\vspace*{-1.3cm}

\begin{IEEEbiography}[{\includegraphics[width=1in,height=1.25in,clip,keepaspectratio]{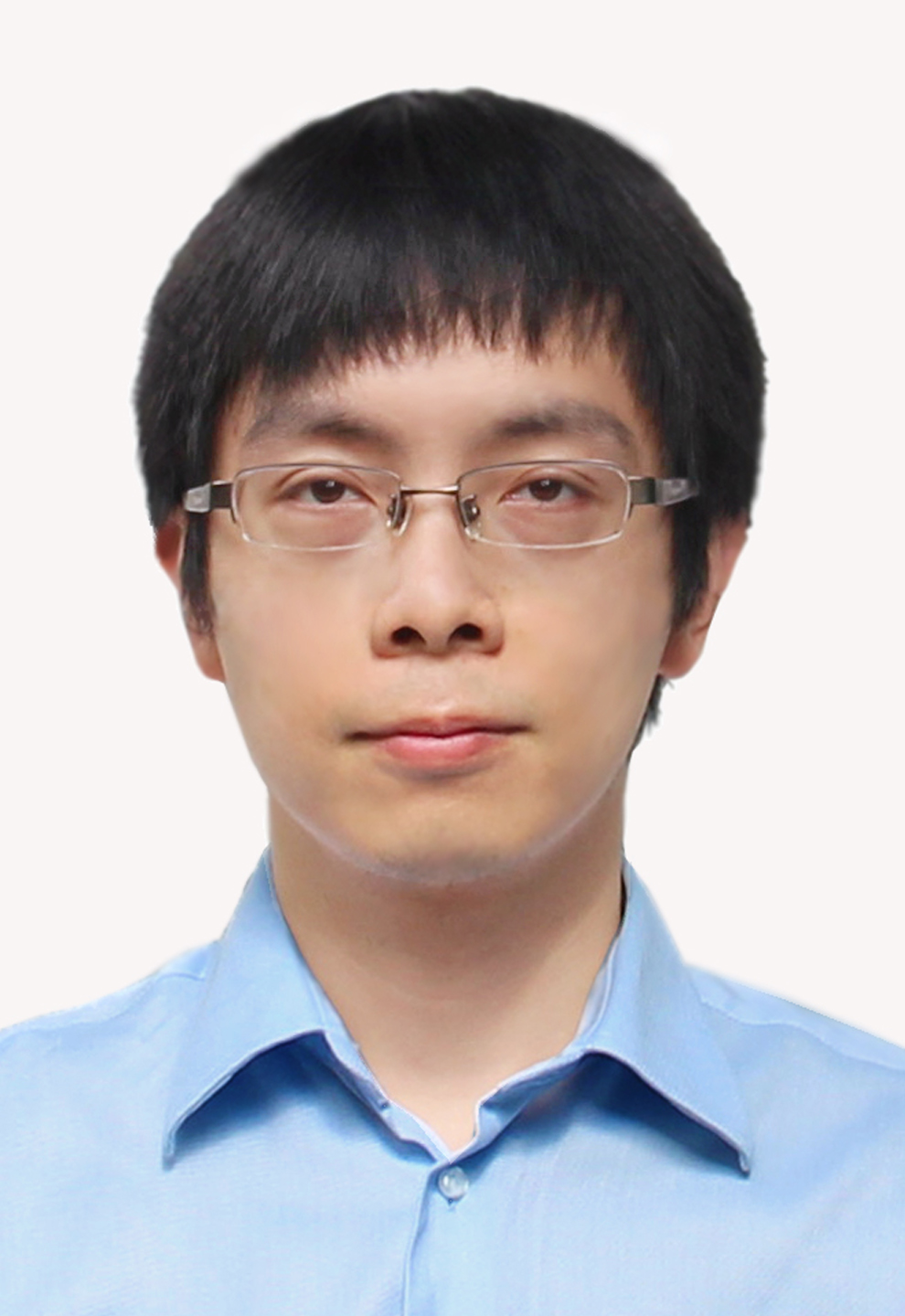}}]{Jiajie Xu}
received the MS degree from The University of Queensland in 2006 and the PhD degree from the Swinburne University of Technology in 2011. He is currently an Associate Professor with the School of Computer Science and Technology, Soochow University, China. His research interests include spatiotemporal database systems, big data analytics, mobile computing, and recommendation systems.
\end{IEEEbiography}

\vspace*{-1.3cm}

\begin{IEEEbiography}[{\includegraphics[width=1in,height=1.25in,clip,keepaspectratio]{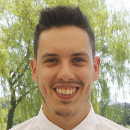}}]{Matteo Francia}
is a PhD candidate in Computer Science at The University of Bologna, Italy. He was a visiting scholar at The University of Queensland in 2019. He received the MSc and BSc with honors from the University of Bologna in 2017 and 2014, respectively. His research focuses on analytics of unconventional data, with particular reference to trajectory, social, and sensory data.
\end{IEEEbiography}

\vspace*{-1.3cm}

\begin{IEEEbiography}[{\includegraphics[width=1in,height=1.25in,clip,keepaspectratio]{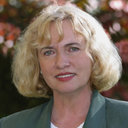}}]{Maria E Orlowska}
is a Professor at Polish-Japanese Academy of Information Technology in Warsaw, Poland. She was Professor of Information Systems at The University of Queensland from 1988 to 2016. She is a Fellow of the Australian Academy of Science. Her main research interests include databases and business IT systems with a focus on modeling and enforcement issues of business processes.
\end{IEEEbiography}

\vspace*{-1.3cm}

\begin{IEEEbiography}[{\includegraphics[width=1in,height=1.25in,clip,keepaspectratio]{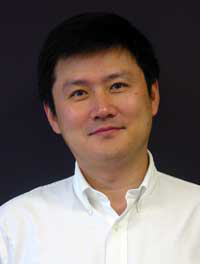}}]{Xiaofang Zhou}
	 is a Professor of Computer Science at The University of Queensland. He received his BSc and MSc in Computer Science degrees from Nanjing University in 1984 and 1987,respectively, and his PhD in Computer Science from UQ in 1994. His research interests include spatial and multimedia databases, high performance query processing, data mining, data quality management, and machine learning. He is a Fellow of IEEE.
\end{IEEEbiography}

\end{document}